%% file: main.tex
\documentclass[acmsmall,screen]{acmart}

\usepackage[inline]{enumitem}
\usepackage{listings}
\usepackage{graphicx}
\usepackage{subcaption, caption}
\usepackage{pagecolor}
\usepackage{xcolor}
\usepackage{soul}
\usepackage{svg}
\usepackage{amsmath}
\usepackage{caption}
\usepackage[linesnumbered,vlined,figure]{algorithm2e}
\usepackage{multicol}
\usepackage{mathtools}
\usepackage{booktabs}
\usepackage{multirow}
\usepackage{natbib}
\usepackage{pifont}
\usepackage{xcolor}

\input{notations.tex}

\extendedAbsract{
}{
\settopmatter{printacmref=false}
}

\setcopyright{rightsretained}
\extendedAbsract{
\acmDOI{10.1145/3689803}
\acmYear{2024}
\acmJournal{PACMPL}
\acmVolume{8}
\acmNumber{OOPSLA2}
\acmArticle{363}
\acmMonth{10}
\acmSubmissionID{oopslab24main-p1030-p}
\received{2024-04-06}
\received[accepted]{2024-08-18}
}{}

\ccsdesc[500]{Theory of computation~Program analysis}
\ccsdesc[500]{Software and its engineering~Just-in-time compilers}
\ccsdesc[300]{Software and its engineering~Dynamic analysis}

\keywords{Java program analysis, staged analysis, 
    points-to analysis, IDFA}

\begin{document}
\title{The ART of Sharing Points-to Analysis\extendedAbsract{}{ (Extended
Abstract)}}
\subtitle{Reusing Points-to Analysis Results Safely and Efficiently}

\author{Shashin Halalingaiah}
\orcid{0000-0002-1268-4345}
\affiliation{%
  \institution{Dept of CSE, IIT Madras}
  \city{Chennai}
  \country{India}
}
\affiliation{%
  \institution{University of Texas at Austin}
  \city{Austin}
  \country{USA}
}
\email{shashin@cs.utexas.edu}

\author{Vijay Sundaresan}
\orcid{0009-0006-9342-4356}
\affiliation{%
  \institution{IBM Canada Lab}
  \city{Markham}
  \country{Canada}
}
\email{vijaysun@ca.ibm.com}

\author{Daryl Maier}
\orcid{0009-0006-8467-3270}
\affiliation{%
  \institution{IBM Canada Lab}
  \city{Markham}
  \country{Canada}
}
\email{maier@ca.ibm.com}

\author{V. Krishna Nandivada}
\orcid{0000-0002-5949-0046}
\affiliation{%
  \institution{Dept of CSE, IIT Madras}
  \city{Chennai}
  \country{India}
}
\email{nvk@iitm.ac.in}

\begin{abstract}

\input{abstract}
\end{abstract}
\maketitle{}
\newpage
\input{intro}

\input{back}

\input{art}

\input{opt}
\input{discuss}

\input{impl}
\input{eval}

\input{related}
\input{concl}

\bibliographystyle{ACM-Reference-Format}
\bibliography{refs}
\end{document}

%% file: notations.tex
\newcommand{\iloop}{\mathcal{I}_{loop}}
\newcommand{\iout}{\mathcal{I}_{out}}
\newcommand{\iin}{\mathcal{I}_{in}}

\newcommand{\out}[1]{\text{\rm OUT}[#1]}
\newcommand{\subsumes}{\succcurlyeq}
\newcommand{\arti}[1]{\ensuremath{ART\langle#1\rangle}}
\def\projectin{\mbox{\tt project-in}}
\def\artwork{\mbox{\rm {\sc art}work}}
\def\artExpanded{\mbox{Analysis-results Representation Template}}
\newcommand{\checkIntra}[1]{{\tt checkForIntraSafety}(#1)}
\newcommand{\checkOut}[1]{{\tt checkForOutSafety}(#1)}
\newcommand{\checkIn}[1]{{\tt checkForInSafety}(#1)}
\newcommand{\extendedAbsract}[2]{#2} 

\def\regenpta{\mbox{\rm RegenPTA}}

%% file: abstract.tex
Data-flow analyses like points-to analysis can vastly improve the precision of other analyses, and
enable powerful code optimizations.
However, whole-program points-to analysis of large Java programs tends to be expensive -- both in terms of time and memory.
Consequently, many compilers (both static and JIT) and program-analysis
tools tend to employ faster -- but more conservative -- points-to analyses to improve usability.
As an alternative to such trading of precision for performance,
various techniques have been proposed to perform precise yet expensive
fixed-point points-to analyses ahead of time in a static analyzer, store the
results, and then transmit them to independent
compilation/program-analysis stages that may need them.
However, an underlying concern of safety affects all such techniques --
can a compiler (or program analysis tool) trust the 
points-to analysis results generated by another compiler/tool?

In this work,  we address this issue of trust in the context of Java,  while accounting for the issue of performance.
We propose \textbf{ART}: \artExpanded{} – a novel scheme to efficiently
and concisely encode results of flow-sensitive, context-insensitive points-to analysis
computed by a static analyzer for use in any independent system that may
benefit from such a  precise points-to analysis.
ART also allows for fast regeneration of the encoded sound analysis results in such systems.
Our scheme has two components: (i) a producer that can statically perform expensive
points-to analysis and encode the same concisely, (ii) a consumer that, on
receiving such encoded results (called \artwork{}), can regenerate the
points-to analysis results encoded by the \artwork{} if it is deemed
``safe''.
The regeneration scheme completely avoids fixed-point computations and thus can help consumers like static analyzers and JIT compilers to obtain precise points-to information without paying a prohibitively high cost.
We demonstrate the usage of ART by implementing a producer (in Soot)
and two consumers (in Soot and the Eclipse OpenJ9 JIT compiler).
We have evaluated our implementation over various benchmarks from the DaCapo and SPECjvm2008 suites.
Our results demonstrate that using ART, a consumer can obtain precise
flow-sensitive, context-insensitive points-to analysis results 
in less than (average) 1\% of the time taken by a static analyzer
to perform the same analysis, with the storage overhead of ART
representing a small fraction of the program size (average around 4\%).

%% file: intro.tex
\section{Introduction}
\label{s:intro}

One of the salient features of Java  is portability.
Programs written in Java are compiled once by their respective
static compilers to a platform-independent intermediate language (bytecode). 
Once statically compiled, these programs may then be analyzed by
independent program analysis tools, or
executed
on platform-specific Java Virtual Machines (JVMs)
that
use just-in-time (JIT) compilers to generate optimized native code specific to the target platform.
Such two-step compilation processes bring in unique opportunities and interesting challenges.

For example,
the time spent in JIT compilation is considered a part of the execution time of the program as a whole, as the JIT compilation is performed at runtime.
Thus, it is essential that the time spent in JIT compilation is not
prohibitively high as this could render such JIT compilers unusable in practice.
A direct impact of such a restriction is that 
all the JIT compilers in popular Java virtual machines (like Eclipse OpenJ9~\cite{openj9}, the HotSpot JVM~\cite{10.5555/1267847.1267848}, Jikes RVM~\cite{5386722}, among others) 
avoid complex fixed-point-based iterative data-flow analyses (for example,
inter-procedural points-to analysis) and instead
utilize imprecise analyses in the form of approximations and heuristics.
Any optimizations performed by the JIT compilers based on such imprecise
points-to analysis results tend to be less effective.
In summary, these JIT compilers
end up sacrificing analysis precision for execution efficiency.

There have been prior works that
make precise points-to analyses results available to the JIT compiler without
negatively impacting its performance, 
by using various staged analysis~\cite{ali_2014,serrano,10.5555/902440,pye}.
In most such techniques, a producer (static compilation) stage bears the cost of
expensive points-to analysis and makes the results of the analysis available to the
consumer  (JIT compiler); this information may even be transmitted over the wire.
The consumer, in turn, simply reads the obtained results, and uses
them when needed after making any necessary (non-expensive) adaptations.
However, owing to the remote and independent nature of the consumer, two
important challenges exist:\\
\noindent
1.  {\em Ensuring safety} -- given that the virtual machine is an independent (and
    possibly remote) system, how does it ensure that the results of an
    obtained expensive points-to analysis are safe to use? Note that
    the analysis results may have been tampered  by a malicious producer.
Since the requirement of sound analysis results is non-negotiable for
JIT compilers, 
production JVMs 
cannot ``just trust'' the static analysis results as they are. 
 Since there is no existing way for such a consumer to trust the analysis
 results, 
at this time, it is not common in practice for a production JIT compiler to
  consume static analysis results from outside.
Further,  if such a JIT compiler wants to consume externally-sourced static analysis
results, then having some sort of inexpensive verification scheme is essential. 


\noindent
 2.  {\em Ensuring transmission efficiency} -- for a large program
    (where an expensive points-to analysis has the most benefit), the artifacts of
    an analysis can potentially be huge.
    Since any such artifacts have to be transmitted to the JVM along with the bytecode, it effectively increases the size of the executable and is thus considered an overhead that impacts portability.
    Further, reading large-sized artifacts may impact the performance of
    the JIT compiler negatively.

    Similar to the impact on JIT compilers, the precision of many useful
    static analyses and optimizations (for example, Function
    Inlining~\cite{10.5555/646153.679523}, Stack
    Allocation~\cite{10.1145/320385.320400}, Common Sub-expression
    Elimination~\cite{muchnick}, and so on) 
    can be directly improved by employing more precise
    points-to analyses.
    In a typical static analysis workflow, points-to analysis may need to be run on a program as a pre-pass.
    Attempting to use a precise fixed-point-based points-to analysis in
    each of these analyses may result in an unacceptable analysis-time overhead.
    Inspired by the staged compilation discussion above, one can envisage
    a scenario where a producer performs an expensive analysis once and
    stores the results in a persistent store. 
    These results may be read  by later consumers (independent static analysis/compiler tools) to
    obtain/use the analysis results efficiently.
    However, the same concerns of safety and transmission efficiency hold in this scenario as well.

\begin{figure}[t]
	\small
	\begin{minipage}{0.03\textwidth}
		~
	\end{minipage}
	\begin{minipage}{0.42\textwidth}
\begin{lstlisting}[]
void foo() {
	A a = new A();$\label{ln:moti-ex-allocation-var}$
	a.f = new F1();$\label{ln:moti-ex-allocation-heap}$
	A b = new B();
	b.f = new F2();
	A c = new C();
	c.f = new F2();
	while(*)         {$\label{ln:moti-ex-loop-start}$ // unknown condition.
		c.f = b.f;
		b.f = a.f; $\label{ln:moti-ex-loop-end}$ } // end loop
	... $\label{ln:post-loop}$ } // end foo
\end{lstlisting}
	\caption{A synthetic Java program whose points-to analysis necessitates
	fixed-point computation due to the presence of a loop.}
	\label{fig:loop-motivating}
	\end{minipage}
	\begin{minipage}{0.03\textwidth}
		~
	\end{minipage}
	\vline
	\begin{minipage}{0.03\textwidth}
		~
	\end{minipage}
	\begin{minipage}{0.46\textwidth}
  \includegraphics{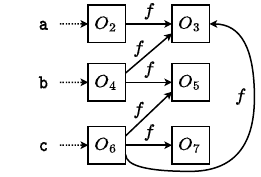}
  ~\\~\\~\\~\\
  	\caption{Points-to graph representing the fixed-point points-to information 
		at Line~\ref{ln:moti-ex-loop-start} of the program snippet in
		Fig.~\ref{fig:loop-motivating}.
		}
	\label{fig:art-intra-iloop}
	\end{minipage}
\end{figure}

We now use an example to illustrate these challenges by using an
optimizing compiler as a consumer.
Consider the Java program snippet shown in Fig.~\ref{fig:loop-motivating}.
Analyzing the code manually, we can see that after the loop
body is executed, at Line~\ref{ln:post-loop}, the field {\tt c.f} may
point-to one of the three different objects (of two different types -- {\tt F1} or {\tt F2}).
Determining this fact precisely requires  precise points-to analysis
results, which in turn requires expensive fixed-point computations.
To avoid such performance overheads in the consumer, say we use staged analysis:
we can use a static analyzer to perform expensive points-to analysis (for
example, flow-sensitive) and transmit the results to the consumer.
Such a flow-sensitive analysis would rightly indicate that at Line~\ref{ln:post-loop}, {\tt c.f} may point to objects of two
types.
Say, these analysis results were tampered with to indicate that at
Line~\ref{ln:post-loop}, {\tt c.f} points to a single object.
If the consumer uses this tampered information, then it may
lead to optimizations or analysis-results that are not semantics-preserving.
For example, if there is a call statement {\tt c.f.bar()} after
Line~\ref{ln:post-loop}, then the call may be erroneously inlined by the
function inliner of the optimizing compiler.

We can also see the challenges related to transmission efficiency:
a naive scheme that supports flow-sensitive points-to
analysis, and transmits the flow-sensitive information at each
program-point, will lead to potentially high space and time overheads.
These issues of safety and transmission efficiency can also be observed
in programs with function calls and recursion.

In the context of inter-procedural analysis, considering the scalability (space and time) issues arising in
context-sensitive analysis~\cite{SmaragdakisBalatsouras15}, context-insensitive
analysis provides a middle ground, compared to a fully intra-procedural
analysis.
In this article, we propose a scheme to encode 
context-insensitive points-to analysis results for Java programs to enable
inter-stage transmission in a manner that addresses both the issues
discussed above.
We add flow-sensitivity to maintain increased precision~\cite{pye,ors} without much
performance trade-offs.
We term our proposed scheme ART (\artExpanded{}).

At its core, given a flow-sensitive, context-insensitive points-to analysis result generated by a producer (say, a static analyzer), 
ART prescribes a small subset of representative information that is enough
to encode the results of the whole analysis for efficient transmission to
a consumer. 
We call this encoded information the \artwork{} for the given analysis and program.
In addition, the information selected for encoding by ART is such that it
allows the consumer to not only establish the safety of the received
\artwork{}, but also very efficiently regenerate the complete points-to
analysis results encoded by this received \artwork{}, if found safe.

There have been works that augment code in ways that enable verification of the code's safety during execution.
TAL (Typed Assembly Language~\cite{tal}) is one such example where assembly code is type-checked to certify the code, which can be used in systems where code must be checked (before execution) for untrusted and potentially malicious behavior.
Similarly, PCC (Proof-carrying code~\cite{necula}) is a novel mechanism where a host system can ascertain whether it is safe to execute a program provided by an untrusted source.
In PCC, the producer of the untrusted code must supply -- along with the code -- a safety proof that attests to the code's adherence to a previously agreed upon safety policy.
These works have led to further research targeting different languages and domains~\cite{10.1007/978-3-662-58820-8_22,10.1007/978-3-030-05998-9_13,10.1145/3477579}.
Starting from Java 6, the JVMs support StackMapTable~\cite{jvms} attributes, using which the static compiler can communicate type information to JVMs, which can be modularly verified and used to perform type checking of Java applications.
Our technique of ART is similar in spirit, wherein ART uses
a fundamental property of the underlying (points-to) analysis to establish its safety at the consumer end, but novel in that it applies the safety reasoning to points-to analysis results.
To the best of our knowledge there is no prior work that caters to sharing of points-to analysis results in a safe and efficient manner.

The design of ART is based on the underlying properties of fixed-point
computations.
Given an \artwork{}, for an analysis $A$ and program $P$, the consumer
makes exactly one pass over the statements of the whole-program and
regenerates the complete points-to analysis results at each program-point by using
the information present in the \artwork{}, thereby avoiding any fixed-point computation.
In the process, it establishes the safety of the \artwork{}, by checking
that the received \artwork{} is consistent with the results regenerated
by the consumer.
We define the notion of safety for any \artwork{} and
give a guarantee that for a given program $P$ and a points-to analysis $A$, the results regenerated from a safe
\artwork{} matches the underlying fixed-point results.


To summarize the different approaches discussed above:
compared to the performance-centric systems (that avoid expensive
fixed-point computation based analyses) and precision-centric systems (that
perform expensive fixed-point computations to ensure high precision), the
staged analysis supported systems discussed before~\cite{ali_2014,serrano,10.5555/902440,pye} provide both precision and efficiency,
but at the cost of soundness.
In contrast, ART helps us obtain highly
precise points-to results, without performing any fixed-point computation, while
guaranteeing the soundness of the obtained results.

In this paper, we describe ART for flow-sensitive, context-insensitive
points-to analysis of Java programs. 
However, we believe that the discussed concepts can be extended to flow-sensitive, context-sensitive points-to analysis, as well as, any flow-sensitive,
context-(in)sensitive iterative data-flow analysis, for any Java-like languages.

\noindent
{\bf Contributions}\\
~
$\bullet$
We propose ART, an efficient scheme for a producer to concisely encode the
results of whole program flow-sensitive, context-insensitive points-to
analysis, for Java programs; this encoded result is termed \artwork{}.  \\
$\bullet$	  
We propose a scheme that a consumer may employ to efficiently regenerate
the originally encoded whole-program points-to analysis results from the
given \artwork{}, while ensuring safety. Our scheme is accompanied by a
correctness argument.\\
$\bullet$	  We demonstrate the usage of ART by instantiating 
 two consumers (one in the Soot framework, and one in the Eclipse OpenJ9 JIT
 compiler) that share a common producer (implemented in Soot). \\
%
%
\noindent
$\bullet$
We evaluate our implementation over various benchmarks from the DaCapo and SPECjvm2008 suites.
    The evaluation shows that ART enables a consumer to obtain and gainfully use
    previously unattainable precise flow-sensitive points-to analysis results
    without making the consumer's execution time prohibitively high, while addressing
    the issues of safety, and transmission efficiency.
%

%
%


%% file: back.tex
\section{Background}
\label{s:back}
In this section, we provide a brief description of some background
material relevant to this paper.

\label{ss:back-prog-rep}
{\em Recursive call-site}. Consider a call-graph with cycles (due to
recursion). A call-site that corresponds to an edge in the
cycle of a call-graph, is a recursive call-site.
For example, if {\tt
main} calls {\tt foo}, {\tt foo} calls {\tt bar}, and {\tt bar} calls {\tt
foo}. 
The call-sites of the latter two calls are considered recursive call-sites.

%

{\em Points-to Analysis.}
\label{ss:back-pta}
Points-to analysis is a program-analysis technique that can establish which storage locations (or objects) are pointed to by which pointers (or reference variables).
We use a points-to graph (similar to \citet{pye}) to represent the points
to information.
A points-to graph $G(V, E)$ consists of
(i)~a set $V$ of nodes representing the variables and abstract objects in the
program; and
(ii)~a set $E\subseteq (V\times V) \cup (V \times {\tt Fields} \times V)$ of edges representing the points-to relationships among the
nodes in the program.
An edge $(a, O_x)$ from a reference variable $a$ to a node $O_x$ in a
points-to graph implies that the variable $a$ may point to the object $O_x$.
Similarly, an edge~$(O_x, f, O_y)$ from node $O_x$ to $O_y$ with a
label $f$ implies that $O_x.f$ may point to $O_y$.
Statically, we represent an object on the heap by an abstract-object $O_l$ where $l$ is a label indicating the line number of the program where the object is allocated.

In this paper, we use a pictorial representation of points-to graphs to illustrate points-to information at various program statements.
In a points-to graph, we use dotted lines to represent edges between a reference variable and a heap object, and solid lines to represent edges between heap objects.
%
For example, Fig.~\ref{fig:art-intra-iloop} shows the points-to graph 
at Line~\ref{ln:moti-ex-loop-start} of the program-snippet shown in
Fig.~\ref{fig:loop-motivating}, after completing the fixed-point
points-to analysis.


{\em Iterative Data Flow Analysis (IDFA).} An iterative data-flow analysis is
defined over lattice $\mathcal{L}$, and a set of flow-equations (or,
transfer functions) that establishes the relationship between data-flow
values.
Flow-sensitive context-insensitive points-to analysis is typically encoded as an IDFA, where the goal is to
compute the points-to graph after each statement.
At each iteration of analysis, the information flowing {\em in} to a node
(called the IN-value of that node) is transformed, by applying the
flow-equation of the node, into information flowing {\em out} of it (called the OUT-value of that node).
This evaluation of flow-equations continues in an iterative manner until the points-to information at each node stabilizes (that is, reaches a {\em fixed-point}).
For a program $P$,
we say that the result $R$ of a points-to analysis $A$ is the fixed-point result for $A$, if $R$ satisfies all the flow-equations of $A$.
Naturally, given an analysis $A$ and a program $P$, there may be many fixed-point results for $P$ that individually satisfy all the flow-equations of $A$.
However, $P$ will have a single {\em least fixed-point result} for $A$.
In the context of points-to analysis in this paper, the lattice is the
power-set of abstract objects, the $\bot$ element is represented by the
set of all the abstract objects,
$\top$ is represented by the empty set, and the meet operator is
given by the set union operator.

In context-insensitive analysis, the summary of points-to information
flowing into a procedure is known as the IN-flow (or IN-summary) of the
procedure; and upon completion of analysis of a procedure, the summary of
the information flowing out of it back to the call-site is called its OUT-flow (or OUT-summary).
At a call-site, computing the information flowing in to a callee varies based on the underlying analysis.
In general, it involves taking a projection of the IN-value of the
call-site with respect to the portion of heap reachable from the
actual arguments of the function call.
We encode this process by a macro {\tt project-in}.
Similarly, we use a macro {\tt project-out} to propagate the points-to
information from the {\tt Exit} node of any procedure  back to the call-site.
We note that the specific mechanism of doing this does not weigh in on our technique, and so we do not delve into it in this paper.


{\em Comparison of points-to information.}
\label{ss:back-subsumes}
Given two instances of points-to information $I_1$ and $I_2$ represented by points-to graphs $G_1$ and $G_2$ respectively, we say: 
	 that $I_1$ is ``equal to'' or ``matches'' $I_2$ if $G_1$ = $G_2$. 
 	Similarly, we say that $I_1$ ``subsumes'' $I_2$ (represented as $I_1 \subsumes I_2$) if $G_2$ is a subgraph of $G_1$.

%% file: art.tex
\section{ART: \artExpanded{} for Java Programs}
\label{l:art}
\label{s:art}
%

As discussed in Section~\ref{s:intro},
since the producer and the consumer of points-to analysis results can be independent systems, 
the consumer needs a way to check that the obtained analysis results are sound with respect to
the program under consideration.
In this section, we discuss an encoding scheme called ART to (i) efficiently encode a summary of the points-to analysis
computed by a producer, and (ii) quickly regenerate the sound analysis
results represented by the encoding (if the encoding is found to be safe) in a consumer.
In this scheme,
the encoding is efficient in that it 
avoids sending the complete analysis results, but 
includes minimal information 
that can be used by the consumer to regenerate the analysis results (termed as the {\em efficiency-goal of the
producer}).
Similarly, at the consumer site, the analysis results are generated quickly in the
sense that it does not require any fixed-point computation (called the
{\em efficiency-goal of the consumer}).
We term any given instance of ART for a program $P$ and analysis $A$ as the
\artwork{} for $P$ and $A$.

Before we present our proposed scheme, 
we first state our basic assumptions about the producer and the consumer:
For a given program $P$ and a fixed-point computation based points-to analysis $A$,
(i) the consumer needs the results for $A$,
but cannot afford to compute it from scratch,
(ii) the producer makes available an \artwork{} purported to be that for $P$
and $A$, and
(iii) the consumer can access that \artwork{}.
In Section~\ref{l:discuss}, we discuss a relaxation of this assumption,
where the producer and consumer may not refer to the same points-to analysis.
%
%
Further, the consumer expects the following two soundness and completeness
guarantees: (i) unsound analysis results will never be marked as sound,
and (ii) sound analysis results will always be marked as sound.

For ease of exposition,
we will first discuss ART in the context of intra-procedural
flow-sensitive points-to analysis and 
then extend it to handle inter-procedural points-to analysis (in
Section~\ref{ss:art-inter-design}).
In this paper, we will use a notion of safety given by the following
definition.
\begin{definition}
Given an intra (inter) procedural iterative-data-flow points-to analysis $A$, and a program $P$,
we say that an intra (inter)  procedural \artwork{} is {\em safe} for $P$, with respect to $A$,  
if it encodes a sound intra (inter) procedural points-to analysis result of $P$
satisfying the transfer functions of $A$.
\label{def:safety}
\end{definition}

\input{art-intra-design}
\input{art-intra-regen}
\input{art-intra-safety}

\input{art-inter-design}

\input{art-inter-regen}
\input{art-inter-safety}
\input{art-correctness}

%% file: art-intra-design.tex
\subsection{Design of Intra-procedural ART}
\label{s:art-intra-design}
\label{ss:art-intra-design}
During intra-procedural flow-sensitive points-to analysis, 
it is well understood that computing the points-to information for 
statements inside loops involve fixed-point computation and hence is
more expensive (compared to statements outside the loops).
On the other hand,
for a statement that is not inside a loop, its OUT can be computed, without needing any fixed-point computation, if we have the OUT of its topologically sorted predecessors.
We use this understanding to design ART.


The intuition behind the design of ART is that it should only carry information
that would otherwise be expensive to compute, and
the consumer must be able to use this information to regenerate the encoded analysis.
To understand the minimal information that needs to be included in ART,
consider the results realized at the end of a flow-sensitive
points-to analysis for a procedure. 
These results  include the OUT for each
statement.
Say we have the control-flow graph (CFG)~\cite{muchnick} of the procedure and
the OUT of the {\tt Entry} node of the procedure is also given.
%
%
The first instruction of a basic-block is termed a leader~\cite{muchnick}.
Consider a basic-block, whose leader is a loop-header.
We can avoid storing the point-to information of all the statements
in the basic-block and recompute them without needing any fixed-point computations, 
if we have the fixed-point OUT of the loop-header.
We term such a basic-block whose leader is an {\tt Entry} node or a loop-header as a
{\em key} basic-block.
For the non-key basic-blocks the IN points-to information of their leaders can
be computed simply by taking the meet of the OUTs of their respective
predecessors. 
This IN information can be used to compute the OUT values of
the leaders of these non-key basic-blocks;
thus the OUT values for such leaders or the statements in their
corresponding basic-blocks need not be stored.

To summarize, the OUT information 
of the leaders of the key basic-blocks can be used to generate the points-to
information for all the statements, without needing any fixed-point
computation (the exact scheme of regeneration will be discussed in
Section~\ref{ss:art-intra-regen}).

Since OUT of the {\tt Entry} node is
equal to its IN, which in intra-procedural analysis is initialized to $\bot$
(see Section~\ref{s:back}) it need not be stored.
In Section~\ref{ss:art-inter-design}, we will revisit this point when we discuss
how ART deals with inter-procedural analysis.

We use the discussion above to identify the single type of points-to information that ART needs to carry in order to encode intra-procedural points-to analysis:

\noindent $\bullet$ \textbf{Loop-Invariants:}
    A {loop-invariant} encodes the fixed-point OUT of a loop-header. 
    We denote the collection of all such loop-invariants
using a map $\iloop : L \rightarrow G$, where $L$ is the set of labels of all the
loop-headers  and $G$ is the set of all possible OUTs.


{\bf Example.} For the program snippet shown in
Fig.~\ref{fig:loop-motivating}, ART would need to encode only the
OUT values corresponding to the loop-header $s_{\ref{ln:moti-ex-loop-start}}$. 
For each program $P$, the list of such encodings (program-points and the
corresponding OUT values) is called an instance of ART, represented as
\arti{P}. We use $\arti{P}.\iloop$ to refer to the loop-invariant map of
\arti{P}.
The contents of $\arti{P}.\iloop[s_{\ref{ln:moti-ex-loop-start}}]$ for the program snippet in Fig.~\ref{fig:loop-motivating} are shown in Fig.~\ref{fig:art-intra-iloop}.

We would like to highlight that 
while the traditional data-flow analyses~\cite{muchnick} generate the most precise
fixed-point result, in general,
there can be many fixed-point results (and hence loop-invariants) for a
given loop. 
And each of them can be a valid instance of ART encoding for that
loop.
Thus a program can have many valid instances of ART, each of which
encodes/corresponds to a sound point-to analysis result for that program.

%% file: art-intra-regen.tex
\subsection{Regeneration Of Sound Intra-procedural Analysis}
\label{ss:art-intra-regen}

We will now present a technique to regenerate the sound intra-procedural
points-to analysis for a program $P$, using the given instance \arti{P},
such that the generated points-to results match the points-to
results encoded by \arti{P}.
For ease of understanding, in this section, we assume that the 
\arti{P} is safe (Definition~\ref{def:safety}).
In Section~\ref{ss:art-intra-safety}, 
we discuss how to verify the safety of the given \arti{P}.

Fig.~\ref{fig:art-intra-regen-algo} presents the algorithm for
regeneration of sound intra-procedural points-to analysis for a procedure
$M$ in $P$, given \arti{P};
we use \arti{P,M} to represent the information encoded in \arti{P} for the
program points within $M$.
Our regeneration scheme follows the steps that the producer would have followed to
generate the encoded points-to information, while making sure that we
do not incur any fixed-point computations.


\input{algos/art-intra-regen}

Our scheme ensures that the consumer analyzes each statement of $M$
exactly once.
For an iterative data-flow analysis like points-to analysis, a statement needs to be
reanalyzed only when its IN-value changes during the course of the
analysis (which happens only if the OUT of at least one of its predecessors changes).
We avoid the reanalysis by ensuring that the fixed-point OUTs of all the
predecessors of each statement are obtained before the statement is analyzed.
This required ordering can be ensured by visiting each basic-block 
and each statement within the basic-block in order by ignoring the
back-edges (Loops starting at
Lines~\ref{ln:for-each-ebb} and~\ref{ln:for-each-statement},
Fig.~\ref{fig:art-intra-regen-algo}). 
As discussed in Section~\ref{ss:art-intra-design}
for the Entry node of $M$, the OUT-value is set to $\bot$.
The fixed-point OUTs of the leaders of the key basic-blocks
are obtained from ART (Line~\ref{ln:fixed-point-key-ebb}).
For each such statement $s$, $\out{s}$ can be computed given the OUTs of
all its predecessors and the relevant transfer function $f_s$
(Line~\ref{ln:fixed-point-non-key-ebb}).
Note that since there is an agreement between the producer and the
consumer on the specific points-to analysis to use, both of them use the
same transfer function $f_s$, for each statement $s$.
The method call {\tt checkForIntraSafety} will be used to check the safety of
the obtained ART and is discussed in Section~\ref{ss:art-intra-safety}.




{\bf Complexity.} In our proposed scheme, we process each statement
exactly once, that is $O(N)$, where $N$ is size of the program.
If the cost of any transfer function is $O(g(N))$, then the complexity of
our regeneration scheme is $O(N\times g(N))$.
In contrast, a typical flow-sensitive, context-insensitive
points-to analysis may process the statements up to $O(N^3)$ times and
incurs a cost of $O(N^3 \times g(N))$.

{\bf Example.}
Consider the snippet shown in Fig.~\ref{fig:loop-motivating}.
We use $s_i$ to refer to the CFG node for line $i$ and $f_i$ to denote the transfer function
used by the producer to perform flow-sensitive intra-procedural points-to
analysis, for $s_i$.
The transfer functions used by the producer can be classified into three
categories based on the type of the node:
(i) the {\tt Entry} node: OUT$[{\tt Entry}]=\bot$.
(ii) a node $s_i$ with a single predecessor $p_i$: OUT$[s_i]=f_i(\mbox{OUT}[p_i])$.
(iii) a node $s_i$ with multiple predecessors $\mathcal{P}$:
OUT$[s_i]=f_i(\sqcap_{p\in\mathcal{P}}\mbox{OUT}[p])$.
For example, for the snippet in Fig.~\ref{fig:loop-motivating},
we use the constraint, $\out{s_8} \gets f_8(\out{s_7} \bigsqcap
\out{s_{10}})$.
Note that the constraints to generate OUTs for the statements at
Lines~\ref{ln:moti-ex-loop-start}-\ref{ln:moti-ex-loop-end}
require a fixed-point computation to obtain a solution.


Once the OUTs for all the statements have been computed,
in our suggested scheme the producer will emit 
$\text{OUT}[s_{\ref{ln:moti-ex-loop-start}}]$ as a loop-invariant in
\arti{P,{\tt foo}}, as shown in Fig.~\ref{fig:art-intra-iloop}.
The consumer invokes \regen with the procedure \texttt{foo} and \arti{P, \texttt{foo}} as inputs.


The consumer first initializes $\out{\texttt{entry}}$ to $\bot$.
Except for $\text{OUT}[s_{\ref{ln:moti-ex-loop-start}}]$, 
the constraints to compute the remaining OUTs are exactly the same as that used
by the producer.
For $s_{\ref{ln:moti-ex-loop-start}}$,
the consumer simply uses the constraint:
$\text{OUT}[s_{\ref{ln:moti-ex-loop-start}}] \gets
f_{\ref{ln:moti-ex-loop-start}}(\iloop[s_{\ref{ln:moti-ex-loop-start}}])$.
It can be seen that none of the constraints used by the consumer have cyclic dependencies and
hence need no further fixed-point computation.

Thus, with the assistance of ART,
the consumer has regenerated intra-procedural points-to analysis results
for the whole procedure without needing any fixed-point computations; this
matches the results generated by the producer. 


%% file: algos/art-intra-regen.tex
\SetKwProg{Fn}{Function}{}{end}
\SetKwProg{macro}{Macro}{}{end}
\SetKwFunction{regen}{regenIntra}
\SetKwFunction{nE}{notEmpty}
\SetKwFunction{pop}{pop}
\SetKwFunction{head}{head}
\SetKwFunction{stmts}{statements}
\SetKwFunction{topo}{topologically sorted list of EBBs of}%
\SetKwFunction{cf}{statements of B in control-flow order}%
\SetKwFunction{islead}{is Leader of key EBB}%
\SetKwFunction{getART}{getART}%
\SetKwFunction{get}{get}%
\SetKwFunction{pred}{predOUT}%
\begin{algorithm}[t]
	\small
  \DontPrintSemicolon
  \Fn{\regen{M, \arti{P, M}}} {
    $\textsf{List }\mathcal{B} \gets \textsf{topologically sorted list of
    basic-blocks of } M$\;
    \ForEach{$B \in \mathcal{B}$} {\label{ln:for-each-ebb}
      $\textsf{List }\mathcal{S} \gets \textsf{statements of } B \textsf{ in program order}$\;
      \ForEach{$s \in \mathcal{S}$} {\label{ln:for-each-statement}

	  \lIf{$\text{s}$ is the $\texttt{Entry}$ node of $M$} {
	  $\text{OUT}[s] \gets \text{IN}[s]$; // { = $\bot$}
	  }
	  \lElseIf {$s \textsf{ \upshape is leader of key basic-block}$} {
	    $\text{OUT}[s] \gets \arti{P, M}.\mathcal{I}_{loop}[s]$;\label{ln:fixed-point-key-ebb}
	  }
	  \lElse {
	      	$\text{OUT}[s] \gets f_s(\bigsqcap\limits_{{p \in
		\textsf{preds}(s)}}^{} \text{ OUT}[p])$
		;
		\label{ln:fixed-point-non-key-ebb}
	    }
	    {\tt checkForIntraSafety}($s$, $M$, $OUT$)\;
      }
    }


    \KwRet OUT\;
    }
	\caption{Regeneration of intra-procedural points-to analysis using ART}
	\label{fig:art-intra-regen-algo}
\end{algorithm}

%% file: art-intra-safety.tex
\subsection{Safety of Intra-procedural \textbf{ART}}
\label{ss:art-intra-safety}

The process of regeneration presented in Section~\ref{ss:art-intra-regen}
assumed that \arti{P,M} obtained by the consumer is safe.
We now will discuss the scenarios where \arti{P, M} may have been
{\em tampered} with (by the producer or a malicious middle agent) and how we
can establish the safety of \arti{P, M}.

Without loss of generality, let us assume that \arti{P} is shared via a
file.
Any tampering of the file that breaks the expected syntactic structure is
easily caught.
Hereafter, we only focus on 
that type of tampering which still encodes valid points-to information,
but not the least fixed-point points-to analysis information for $P$.
Thus, we can identify two ways in which \arti{P, M} can be tampered with:
non-conservative and conservative, as described below.
	Let $\alpha = \arti{P}$ be the encoding of the least fixed-point
	points-to analysis results for $P$;
	say the points-to results were obtained using an analysis $A$.\\
$\bullet$	  {\bf non-conservative tampering.}
	Let $\alpha'$ be a modified version of $\alpha$ such that
	$\alpha'$ does not encode any fixed-point result for $A$.
	Then $\alpha'$ no longer encodes any sound
	intra-procedural fixed-point analysis result for $P$.
    In such cases, we say that $\alpha$ has been non-conservatively tampered with, to yield $\alpha'$. 
    A consumer using such a tampered instance of ART in the regeneration
    process will obtain points-to analysis results that
    do not match any fixed-point result for $P$. 
    Thus, by Definition~\ref{def:safety}, any non-conservatively tampered
    ART is unsafe.
    An example of non-conservative tampering would be the removal of a node or
    an edge in a points-to graph that represents the least fixed-point
    value.\\
$\bullet$	{\bf conservative tampering.}  Any other form of tampering is
considered conservative tampering.
    A consumer using such a tampered instance of ART in the regeneration
    process will obtain points-to analysis results that is more conservative than the least
        fixed-point solution, but still sound (that is, an over-approximation).


We first provide a sketch of our proposed scheme to identify if,
for a given procedure $M$ of a program $P$,
\arti{P, M} given to a consumer (to regenerate the points-to analysis
information) is safe. 
Our scheme leverages a fundamental property of fixed-point computations in data-flow analysis.
Consider a loop in the method $M$ of program $P$, where the point-to analysis algorithm of the producer
reached a fixed-point, after processing the loop body $i$ number of times. 
Processing the loop-body $i+k$ ($k \geq 1$) times will not alter the
computed points-to information for any statement in that loop.


We apply a similar reasoning for establishing the safety of \arti{P,M}.
Say, $\arti{P,M}.\iloop[s]$ is the OUT value of a loop-header labeled $s$ encoded in \arti{P, M}.
Note that  $\arti{P,M}.\iloop[s]$ is supposed to be the fixed-point value of $\out{s}$ in
the producer.
Thus, in the consumer, by initializing $\out{s}$ to $\arti{P,M}.\iloop[s]$,
processing each statement of the loop-body once, and eventually re-computing $\out{s}$ after considering all the
predecessors of $s$ (including the ones via the back-edges) should not
change the value of $\out{s}$.



\input{algos/art-intra-safety}

In Fig.~\ref{fig:art-intra-safety-algo},
we use the above intuition to define 
the \checkIntra{} method used in
Fig.~\ref{fig:art-intra-regen-algo}. 
We process
each control-flow successor $s'$, of $s$, such that $(s, s')$ is a
back-edge.
Based on the intuition presented above, we now re-compute OUT of $s'$
considering all the predecessors of $s'$ (which includes the newly
computed $\out{s}$) and compare it with the previously computed OUT of
$s'$ ($prevOut$).
Given that $prevOut$ was initially set to $\arti{P, M}.\iloop[s']$ (which is supposed to be a fixed-point OUT value), it should not change after this re-computation.
We assert the same in Line~\ref{ln:art-intra-safety-subsumes}. 
If this assertion fails, it means that $\arti{P, M}.\iloop[s']$ was not a fixed-point value.
This implies that \arti{P,M} is not safe and does not encode a sound
points-to analysis for procedure $M$, in program $P$.


The discussion of the algorithm in Fig.~\ref{fig:art-intra-regen-algo} so far assumed that for any loop-header $s$, in a
procedure $M$ of program $P$, the $\iloop[s]$ entry exists. 
But in case the tampering involves the deletion of this entry, then the
consumer would not find such an entry in the \artwork{}.
In such a scenario, the consumer infers a default value for 
$\arti{P,M}.\iloop[s]$ (= IN[$s$]), at Line~\ref{ln:fixed-point-key-ebb}; not explicitly shown.
Thus, if the actual fixed-point OUT[$s$] differs from IN[$s$] then
our above-discussed procedure will detect the tampering.
And in case they do not differ, then it implies that despite the omission
the \artwork{} is safe to use; in Section~\ref{s:opt}, we will use this
observation to optimize the encoding of ART.


\begin{figure}[t]
	\small
	\centering
	\hfill
	\begin{subfigure}[b]{0.33\textwidth}
  	\includegraphics{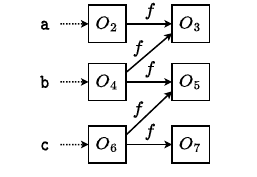}
		\caption{OUT[$s_{\ref{ln:moti-ex-loop-start}}$]$\gets \arti{P, M}.\iloop[s_{\ref{ln:moti-ex-loop-start}}]$ (tampered by
	removing ($O_6, O_3$)).}
	\label{fig:checkIntra-loop-header}
	\end{subfigure}%
	\hfill
	\begin{subfigure}[b]{0.33\textwidth}
  	\includegraphics{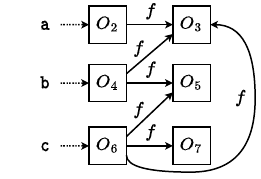}
	~\vspace*{0.1in}
		\caption{Generated OUT[$s_9$], OUT[$s_{10}$].}
	\label{fig:checkIntra-loop-body}
	\end{subfigure}%
	\hfill
	\begin{subfigure}[b]{0.33\textwidth}
  	\includegraphics{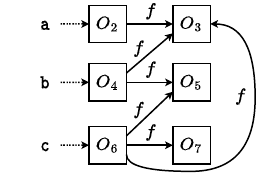}
	~\vspace*{0.1in}
	\caption{Value of $newOut$.}
	\label{fig:checkIntra-newout}
	\end{subfigure}%
	\hfill

	\caption{
	  Values computed by {\tt regenIntra} for the loop-header and the
	  loop-body of Fig.~\ref{fig:loop-motivating}.
	}

\label{fig:art-intra-safety-tampered}
\end{figure}

{\bf Example.}
Let the \artwork{} in Fig.~\ref{fig:art-intra-iloop} be
non-conservatively tampered by removing the edge $(O_6, O_3)$. 
Fig.~\ref{fig:art-intra-safety-tampered} shows
the steps taken by \regen (Fig.~\ref{fig:art-intra-regen-algo}) for the
loop-header and the loop-body in Fig.~\ref{fig:loop-motivating}: 
The algorithm sets $\out{s_{\ref{ln:moti-ex-loop-start}}}$ to the tampered ART
(Fig.~\ref{fig:checkIntra-loop-header}).
Fig.~\ref{fig:checkIntra-loop-body} shows the 
OUT computed for statement $s_9$ (and $s_{10}$).
After computing $\out{s_{10}}$, 
\checkIntra{} (Fig.~\ref{fig:art-intra-safety-algo}) processes the back-edge $(s_{10}, s_8)$ and finds that
the value of $newOut$ (Fig.~\ref{fig:checkIntra-newout}) 
does not match $prevOut$ (Fig.~\ref{fig:checkIntra-loop-header}). 
This leads to the assertion failure at
Line~\ref{ln:art-intra-safety-subsumes},
and establishes that the given \artwork{} has been non-conservatively tampered
with and is thus unsafe to use.

Note: It is simply a coincidence in the above example that $newOut$ matches the non-tampered
\artwork{} (Fig.~\ref{fig:art-intra-iloop}). In general, this may not hold. 
For example, say the tampering removed $(O_4,O_3)$ and $(O_6, O_5)$ edges in addition to
$(O_6, O_3)$, then
$newOut$ will contain  $(O_4, O_3)$ and $(O_6, O_5)$, but not $(O_6, O_3)$.

%% file: algos/art-intra-safety.tex
\SetKwProg{Fn}{Function}{}{end}
\SetKwProg{macro}{Macro}{}{end}
\SetKwFunction{regen}{regenIntra}
\SetKwFunction{nE}{notEmpty}
\SetKwFunction{pop}{pop}
\SetKwFunction{head}{head}
\SetKwFunction{stmts}{statements}
\SetKwFunction{topo}{topologically sorted list of EBBs of}%
\SetKwFunction{cf}{statements of B in control-flow order}%
\SetKwFunction{islead}{is Leader of key EBB}%
\SetKwFunction{getART}{getART}%
\SetKwFunction{get}{get}%
\SetKwFunction{pred}{predOUT}%
\SetKw{assert}{assert}
\SetKw{And}{and}
\SetKw{Or}{or}
\begin{algorithm}[t]
	\small
	\Fn{\checkIntra{s, M, \text{\rm OUT}}} { 
  {
  \DontPrintSemicolon
  \ForEach{$s' \in {succs}(s) $ { such that } $(s, s')$ { is a CFG back-edge} }
 {
     \tcp{$s'$ is a key-basic-block leader and $s'$  has already been analyzed} 
    {
  $prevOut = \out{s'}$  \tcp{was earlier set to $\arti{P, M}.\iloop[s']$}
        $newOut = f_{s'}(\bigsqcap\limits_{p \in \textsf{preds}(s')}^{} \out{p})$\;
	\assert $prevOut = newOut$ \;
	\label{ln:art-intra-safety-subsumes}
 	}
    }
   }
 }
	\caption{Checking Safety of Intra-procedural ART}
	\label{fig:art-intra-safety-algo}
\end{algorithm}

%% file: art-inter-design.tex
\subsection{Design of Inter-procedural ART}
\label{ss:art-inter-design}

We will now extend the intra-procedural ART introduced in
Section~\ref{ss:art-intra-design} to handle inter-procedural
(context-insensitive) points-to analysis.
As discussed in Section~\ref{s:back},
during context-insensitive points-to analysis, a procedure $M$ may be
analyzed more than once till it reaches fixed-point.
Recall that our efficiency-goals (Section~\ref{s:art}) require that we process each statement exactly once  during
the regeneration of the points-to results by the consumer.
By extension, this also requires that we analyze any given
method exactly once, while ensuring that the computed/generated point-to
information at each program-point in that method is valid across all the
calls to that method.
Hereafter, we use points-to IN/OUT information to indicate the
context-insensitive points-to IN/OUT information.

We first present the intuition behind our proposed approach:
Assuming that there is no recursion,
using the intra-procedural ART scheme discussed before,
given the IN-summary of a procedure $M$, we can regenerate the
OUT information for each statement in $M$ (without needing any fixed-point
computation), provided we have the OUT information for each call-site in $M$.
Thus, if we have the IN-summary of every method in the
program and we process the methods of the program, in the bottom-up order of the
call-graph (leaves first), then we can meet our efficiency-goal of the
consumer:
for each method $M$,
the OUT information at each program-point in $M$ (along with the OUT
summary of $M$)
can be computed by processing 
the statements in $M$ exactly once.
But considering the fact that computing the IN-summary for any method $M$
in a context-insensitive analysis involves fixed-point computation, we
propose to encode the IN-summary of each non-recursive method in ART.




The above discussed scheme to efficiently compute the OUT for each
statement of a function given its IN-summary, encounters a challenge
in the case of recursive functions.
This is because
the OUT of the recursive call-sites (see Section~\ref{s:back}) in the recursive methods won't be available
even if we process the call-graph nodes in the bottom-up order.
This challenge can be addressed if the context-insensitive OUT-summary of
each recursive method can be stored (along with the context-insensitive
IN-summary) and made available to the consumer to compute the OUT of the
corresponding recursive call-site.

We use these two intuitions (for recursive and non-recursive procedures)
to propose the addition of the following two types of points-to
information to ART, to support context-insensitive inter-procedural
points-to analysis of any program $P$.

\noindent $\bullet$ {\bf M$_{\bf in}$-Invariants}: An M$_{\rm in}$-Invariant encodes the
		context-insensitive IN-summary of a procedure.
		We denote the collection of all such in-invariants using a
		map $\iin : \mathcal{M} \rightarrow G$, where $\mathcal{M}$ is the set of all
		the procedures in $P$ and $G$ is the set of all possible context-insensitive IN-summaries.

\noindent
        $\bullet$ {\bf M$_{\bf out}$-Invariants}: An M$_{\rm out}$-Invariant encodes the
	  context-insensitive fixed-point OUT-summary of a recursive procedure.
	We denote the collection of all such out-invariants using a map
	$\iout : \mathcal{M}_{rec} \rightarrow G$, where $\mathcal{M}_{rec}$ is the set of all
	recursive procedures (direct and indirect) and G is the set of all
	possible context-insensitive OUT-summaries.

Note that even though both the IN and OUT-summaries of the recursive procedures are
present in the ART of any program $P$, a consumer cannot skip 
processing these recursive procedures during the regeneration process, since it still needs to (i) compute the
OUT for each statement in those methods, and (ii) verify the given IN and
OUT-summaries.


{\bf Summary of the design of ART.} In order to support flow-sensitive, context-insensitive inter-procedural points-to analysis, ART needs to carry three types of points-to information: (a) loop-invariants, (b) in-invariants, (c) out-invariants

\begin{figure}[t]
	\small
	\begin{minipage}{0.03\textwidth}
		~
	\end{minipage}
	\begin{minipage}{0.48\textwidth}
\begin{lstlisting}[]
void main() {
  ...
  A a = new A();
  a.f = new A();
  a.f.f = new A();
  A b = a.f; $\label{ln:moti-ex-foo-main-pred}$
  foo(b, a); $\label{ln:moti-ex-foo-main}$ ...  } $\label{ln:post-recursion}$
void foo(A q, A p) {
  if(p != null) {
    A r = p.f;
    if(*) { p.f = null; } 
    else  { p.f = q; return; }
    foo(q, r); $\label{ln:moti-ex-foo-rec}$
    .../*no effect on heap*/ } }
\end{lstlisting}
	\caption{A synthetic Java program snippet whose points-to analysis necessitates
	fixed-point computation due to the presence of function calls and recursion.}
	\label{fig:rec-motivating}
	\end{minipage}
	\begin{minipage}{0.03\textwidth}
		~
	\end{minipage}
	\begin{minipage}{0.03\textwidth}
		~
	\end{minipage}
	\begin{minipage}{0.38\textwidth}
	\begin{subfigure}{0.45\textwidth}
		\hspace*{-0.3in}
  	\includegraphics[width=2.0\textwidth]{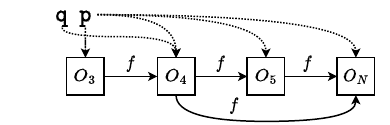}
		\caption{$\iin[{\tt foo}]$}
	\label{fig:i_in_foo}
	\end{subfigure}%

	\begin{subfigure}{0.45\textwidth}
		\begin{center}
  	\includegraphics[width=1.8\textwidth]{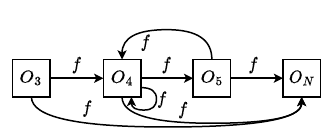}
		\caption{$\iout[{\tt foo}]$}
	\label{fig:i_out_foo}
		\end{center}
	\end{subfigure}%
	\caption{Inter-procedural ART for the program snippet shown in
	Fig.~\ref{fig:rec-motivating}. Here $O_N$ is the abstract object
	representing the {\tt null} object.}
	\label{fig:art-inter-proc}
\end{minipage}
\end{figure}
{\bf Example.}
Fig.~\ref{fig:rec-motivating} shows a program snippet with function calls and recursion.
The contents of $\arti{P}$ for this program snippet 
are shown in Fig.~\ref{fig:art-inter-proc}.
For this program, inter-procedural ART will encode the context-insensitive IN-summary of {\tt foo}.
In addition, since {\tt foo} is recursive, it will also encode the context-insensitive fixed-point OUT-summary of {\tt foo}.

%% file: art-inter-regen.tex
\subsection{Regeneration Of Sound Inter-procedural Analysis}
\label{ss:art-inter-regen}

We will now extend the analysis regeneration scheme discussed in
Section~\ref{ss:art-intra-regen}, to
regenerate the sound inter-procedural points-to analysis for a program $P$, 
using the given instance \arti{P}, 
such that the generated points-to results match the inter-procedural points-to analysis encoded by \arti{P}.
For ease of understanding -- similar to the discussion of intra-procedural regeneration in Section~\ref{ss:art-intra-regen}, we assume that \arti{P} is safe (Definition~\ref{def:safety}).
In Section~\ref{ss:art-inter-safety}, we discuss how to verify the safety of the given inter-procedural \arti{P}.

Fig.~\ref{fig:art-inter-regen-algo} presents the algorithm for regeneration of sound context-insensitive inter-procedural points-to analysis for the whole program $P$ starting from an entry procedure $M$, given \arti{P}.
The presented algorithm builds upon the intra-procedural regeneration
algorithm presented in Fig.~\ref{fig:art-intra-regen-algo} by addressing
regeneration in the presence of method calls. 
We now discuss some of the main differences between the inter-procedural
regeneration scheme and the intra-procedural one.

\input{algos/art-inter-regen}

{\em Handling the {\tt Entry} nodes.}
Since we are now regenerating inter-procedural points-to analysis, 
$\text{IN}$  of the {\tt Entry} node will not be $\bot$;
it is instead initialized with $\arti{P}.\iin[M]$ (Line~\ref{ln:art-inter-entry}).

{\em Handling the call-sites.}
If the statement $s$ is a call-site, we resolve the targets of the
invocation (Line~\ref{ln:art-inter-resolve-targets}).
For each such target $t$, if $t$ has already been analyzed, we do not re-analyze it.
Otherwise, we need to obtain
its OUT-summary so we can compute the OUT information for $s$.
If $s$ is a non-recursive call-site, we recursively invoke \regenInter on $t$ to analyze it and obtain the OUT-summary.
On the other hand, if $s$ is a recursive call-site, computing the OUT-value for $s$ will involve a fixed-point computation.
The consumer avoids this by using the information encoded in \arti{P}, assuming it is safe (Line~\ref{ln:art-inter-recursive}).
After we have obtained the OUT for each target,
we compute $\out{s}$ by taking the meet of the
OUT-summary  of each of the targets
(Line~\ref{ln:art-inter-targets-meet}).
For the rest of the statements, \regenInter follows the same steps as \regen.
The method calls {\tt checkForInSafety} and {\tt checkForOutSafety} will be used to check the safety of
the obtained inter-procedural \artwork{} and are discussed in Section~\ref{ss:art-inter-safety}.

%

{\bf Complexity.} 
The complexity of regeneration of sound inter-procedural
analysis is exactly the same as that of the intra-procedural
analysis (Section~\ref{ss:art-intra-regen}). 





{\bf Example.}
\begin{figure}[t]
	\small
	\centering
	\hfill
	\begin{subfigure}[b]{0.33\textwidth}
  	\includegraphics[width=\textwidth]{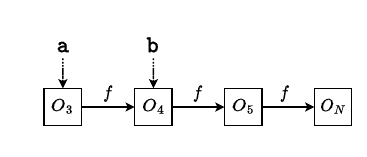}
		\caption{$\text{IN}[s_{\ref{ln:moti-ex-foo-main}}]$  ( $ = \text{OUT}[s_{\ref{ln:moti-ex-foo-main-pred}}]$) }
	\label{fig:art-inter-regen-in-callsite}
	\end{subfigure}%
	\hfill
	\begin{subfigure}[b]{0.33\textwidth}
  	\includegraphics[width=\textwidth]{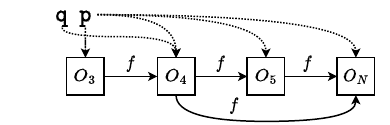}
	\caption{$\text{OUT}[{\tt Entry}]$ of {\tt foo} $(= \iin[{\tt foo}])$}
	\label{fig:art-inter-regen-in-foo}
	\end{subfigure}%
	\hfill
	\begin{subfigure}[b]{0.33\textwidth}
  	\includegraphics[width=\textwidth]{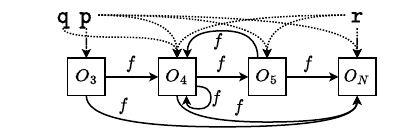}
		\caption{$\text{OUT}[s_{\ref{ln:moti-ex-foo-rec}}]$}
	\label{fig:art-inter-regen-out-rec}
	\end{subfigure}%
	\hfill
	\begin{subfigure}[b]{0.33\textwidth}
  	\includegraphics[width=\textwidth]{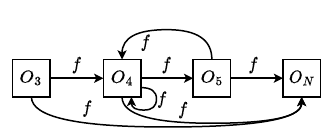}
	\caption{\text{OUT}-summary of \texttt{foo}.}
	\label{fig:art-inter-regen-out-foo}
	\end{subfigure}%
	\hfill
	\begin{subfigure}[b]{0.33\textwidth}
  	\includegraphics[width=\textwidth]{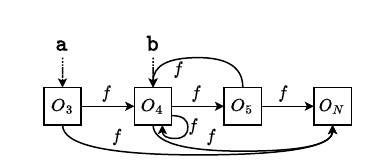}
		\caption{$\text{OUT}[s_{\ref{ln:moti-ex-foo-main}}]$ (computed after analyzing \texttt{foo})}
	\label{fig:art-inter-regen-out-callsite}
	\end{subfigure}%
	\hfill

	\caption{
	  Progression of {\tt regenInter} for program snippet in
	  Fig.~\ref{fig:rec-motivating}.
	}

\label{fig:art-inter-regen}
\end{figure}
For the
program snippet shown in Fig.~\ref{fig:rec-motivating},
given the \artwork{} shown
in Fig.~\ref{fig:art-inter-proc},
in Fig.~\ref{fig:art-inter-regen} we show some of the important steps taken
by \regenInter to regenerate the points-to analysis results; we mainly focus on how we compute the
points-to information in the presence of function calls.
The regeneration process begins at the {\tt main} procedure, and processes
each statement.
Fig.~\ref{fig:art-inter-regen-in-callsite} shows the value of
$\text{IN}[s_{\ref{ln:moti-ex-foo-main}}] (= \text{OUT}[s_{\ref{ln:moti-ex-foo-main-pred}}])$.
To compute $\text{OUT}[s_{\ref{ln:moti-ex-foo-main}}]$, \regenInter needs to process the
function {\tt foo} and obtain its OUT-summary.
To do so, \regenInter first sets the OUT of the {\tt Entry} node of {\tt
foo} to $\iin[{\tt foo}]$ (Fig.~\ref{fig:art-inter-regen-in-foo}),
obtained from the given ART instance (Fig.~\ref{fig:i_in_foo}) and then
processes the statements of {\tt foo}.
At statement $s_{\ref{ln:moti-ex-foo-rec}}$, which is a recursive call,
{\tt regenInter} uses $\iout[{\tt foo}]$ to obtain
the OUT-summary  of ${\tt foo}$ and 
uses it to compute $\text{OUT}[s_{\ref{ln:moti-ex-foo-rec}}]$ (shown in Fig.~\ref{fig:art-inter-regen-out-rec}).
On reaching the {\tt Exit} node of {\tt foo}, {\tt regenInter} will return
the $\text{OUT}$-summary of {\tt foo}
(Fig.~\ref{fig:art-inter-regen-out-foo}) and use it
to compute the $\text{OUT}[s_{\ref{ln:moti-ex-foo-main}}]$
(Fig.~\ref{fig:art-inter-regen-out-callsite}) in the main function.

%% file: algos/art-inter-regen.tex
\SetKwProg{Fn}{Function}{}{end}
\SetKwProg{macro}{Macro}{}{end}
\SetKwFunction{regenInter}{regenInter}
\SetKwFunction{nE}{notEmpty}
\SetKwFunction{pop}{pop}
\SetKwFunction{head}{head}
\SetKwFunction{stmts}{statements}
\SetKwFunction{topo}{topologically sorted list of basic-blocks of}%
\SetKwFunction{cf}{statements of B in control-flow order}%
\SetKwFunction{islead}{is Leader of key EBB}%
\SetKwFunction{getART}{getART}%
\SetKwFunction{get}{get}%
\SetKwFunction{pred}{predOUT}%
\SetKw{And}{and}
\begin{algorithm}[t]
  \DontPrintSemicolon
  \small
  \Fn{\regenInter{M, \arti{P}}} {
    $\textsf{List }\mathcal{B} \gets \textsf{topologically sorted list of basic-blocks of } M$\;
    \ForEach{$B \in \mathcal{B}$} {
      $\textsf{List }\mathcal{S} \gets \textsf{statements of } B \textsf{ in program order}$\;
      \ForEach{$s \in \mathcal{S}$} {
	  \lIf{$\text{s}$ is the $\texttt{Entry}$ node of $M$} {
	  $\text{OUT}[s] \gets \text{IN}[{\tt Entry}]$; // { =  $\arti{P}.\iin[M] $\label{ln:art-inter-entry}}
	  }
	  \lElseIf {$s \textsf{ \upshape is leader of key basic-block}$} {
	    $\text{OUT}[s] \gets \arti{P}.\arti{P, M}.\mathcal{I}_{loop}[s]$;
	  }
	  \ElseIf {$s \textsf{ \upshape is a call-site}$} { 
		$\textsf{List }\mathcal{T} \gets \textsf{ targets resolved for } s$\;\label{ln:art-inter-resolve-targets}
		
		\ForEach{$t \in \mathcal{T}$} {

			  \checkIn{$s$, $t$, \arti{P}, OUT}\;
			  \lIf {$t \textsf{ \upshape has already been
			  analyzed}$} {continue}
			  \If {$s \textsf{ \upshape is a non-recursive
			  invocation of } t$} {
				     $\text{OUT}[t] \gets \regenInter{t, \arti{P}}$\;
			  }
			  \Else ({\em // read from ART to avoid fixed-point computation.}) {
				  $\text{OUT}[t] \gets \arti{P}.\iout[t]$\label{ln:art-inter-recursive}\;\label{ln:art-inter-rec-out}
			  }
		}

		$\text{OUT}[s] \gets \bigsqcap\limits_{t \in
		\mathcal{T}}^{}
		\text{OUT}[t]$\label{ln:art-inter-targets-meet}{\em // As per the underlying points-to-analysis; see Section~\ref{s:back}}\;
	  }
	  \Else {
	      	$\text{OUT}[s] \gets f_s(\bigsqcap\limits_{{p \in
		\textsf{preds}(s)}}^{} \text{ OUT}[p])$
		\;
		\checkIntra{$s$, $M$, OUT}\;
	    }
      }
    }
    \lIf {$M$ is a recursive method} {\checkOut{$M$, OUT};}
    \KwRet OUT\;
    }

	\caption{Regeneration of inter-procedural points-to analysis using ART}
	\label{fig:art-inter-regen-algo}

\end{algorithm}

%% file: art-inter-safety.tex
\subsection{Safety of Inter-procedural ART}
\label{ss:art-inter-safety}

The process of regeneration presented in Section~\ref{ss:art-inter-regen} assumed that \arti{P} obtained by the consumer is safe.
We will now discuss the scenarios where \arti{P} may have been {\em tampered} with and how we can establish the safety of \arti{P}.
Since the two additional points-to information carried by inter-procedural ART (M$_{\bf in}$-Invariants and M$_{\bf out}$-Invariants) are fixed-point values, the discussion on conservative and non-conservative tampering from Section~\ref{ss:art-intra-safety} also holds here.
To recall, any non-conservatively tampered \artwork{} is unsafe.

We note that of the two types of points-to information carried in inter-procedural ART, one of them (M$_{\bf in}$-Invariant) is an IN-value and the other (M$_{\bf out}$-Invariant) is an OUT-value.
As a result, the technique for establishing the safety of each of them has subtle differences -- while still 
using the property of fixed-point values discussed in Section~\ref{ss:art-intra-safety}.
We will now discuss our techniques for establishing the safety of
inter-procedural ART. 

\begin{figure}[t]
\begin{minipage}{0.53\textwidth}
\input{algos/art-inter-safety}
\end{minipage}
~
\begin{minipage}{0.45\textwidth}
	\centering
	\hfill
	\begin{subfigure}[b]{\textwidth}
  	\includegraphics[width=0.75\textwidth]{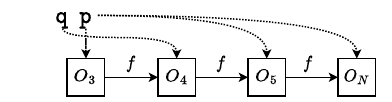}
	\caption{IN[{\tt Entry}]$ \gets \arti{P}.\iin[{\tt foo}]$\\
	(tampered by removing $(p,O_4)$ and $(O_4, O_N)$).}
	\label{fig:art-inter-safety-in-a}
	\end{subfigure}%
	\hfill

	\begin{subfigure}[b]{\textwidth}
  	\includegraphics[width=0.55\textwidth]{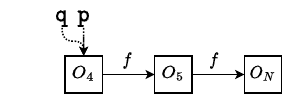}
	~\\
		\caption{computed $in_{\tt foo}$ for the call at $s_{\ref{ln:moti-ex-foo-rec}}$}
	\label{fig:art-inter-safety-in-b}
	\end{subfigure}%
	\hfill

	\caption{
		Safety verification of M$_{\bf in}$-Invariants for the
		procedure {\tt foo}, shown in Fig.~\ref{fig:rec-motivating}.
	}

\label{fig:art-inter-safety-in}
\end{minipage}
\end{figure}

{\bf Establishing Safety of M$_{\bf in}$-Invariants}.
We start with an intuition.
Recall from Section~\ref{ss:art-inter-design} that an M$_{\bf in}$-Invariant encodes the context-insensitive IN-summary of a procedure $M$.
This means that for a procedure $M$, $\iin[M]$ reflects ({\em subsumes}) the IN-summary of $M$ at each and every one of its call-sites.
As a result, processing $M$ by setting the IN of $M$'s {\tt Entry} node to $\iin[M]$ will result in the regeneration of points-to information that is valid at each and every one of $M$'s call-sites.
In contrast, consider a non-conservatively tampered (see
Section~\ref{ss:art-intra-safety}) \arti{P}, where the tampering is performed
with respect to the $\iin[M]$ entry. 
In such a case,
using $\arti{P}.\iin[M]$ (at Line~\ref{ln:art-inter-entry},
Fig.~\ref{fig:art-inter-regen-algo}) 
will result in the generation of points-to information 
(of $M$) that does not correspond to the fixed-point context-insensitive points-to information
of $M$. 
To establish the safety of $\arti{P}.\iin[M]$, we make use of the property of fixed-point values discussed in Section~\ref{ss:art-intra-safety}.
In the producer, if $\arti{P}.\iin[M]$ reached a fixed-point after
analyzing $M$ at all of its call-sites, then processing {\em any} of those
call-sites again in the consumer should not result in any change to the IN-summary of $M$.
We use this intuition to define the {\tt checkForInSafety} method shown in Fig.~\ref{fig:art-inter-safety-algo}.
On encountering a call-site~$s$ invoking a target procedure $T$, {\tt checkForInSafety} first computes $in_T$ by 
invoking {\tt project-in} (see Section~\ref{s:back}) on $\text{IN}[s]$ (Line~\ref{ln:art-inter-safety-project-in}).
It then asserts that the information carried in $\arti{P}.\iin[M]$ {\em subsumes} $in_M$ (Line~\ref{ln:art-inter-safety-in-subsumes}).
This subsumption check is natural (in contrast to an equals check) because
$\arti{P}.\iin[M]$ is supposed to be the meet of the incoming relevant
points-to information from all the call-sites (not just the call-site $s$).
If the assertion succeeds, it means that processing $M$ by using
$\arti{P}.\iin[M]$ in {\tt regenInter} will result in the regeneration of points-to information that is valid at call-site $s$.
If the assertion fails, it means that $\arti{P}.\iin[M]$ does not reflect
the context-insensitive IN-summary of $M$.

The discussion so far assumed that for any procedure $M$ of program $P$,
the $\iin[M]$ entry exists. 
But in case the tampering involves the deletion of this entry, then the
consumer would not find such an entry in the \artwork{}.
In such a scenario, on encountering a call-site $s$ invoking $M$,
{\tt regenInter} sets a default value for 
$\arti{P}.\iin{M}$ (= {\tt project-in}(IN[$s$], $M$)), before the
assertion at Line~\ref{ln:art-inter-safety-in-subsumes},
in Fig.~\ref{fig:art-inter-safety-algo};
not explicitly shown.
Thus, in case the fixed-point context-insensitive OUT-summary of $M$ differs from this default value
then our above discussed procedure will detect the tampering.
And in case they do not differ, then it implies that despite the
omission,
the ART instance is safe to use; in Section~\ref{s:opt}, we will use this
observation to optimize the encoding of ART.

{\bf Example.} 
Consider the \artwork{} shown in Fig.~\ref{fig:art-inter-proc} for 
the program $P$ shown in Fig.~\ref{fig:rec-motivating}.
As an instance of non-conservative tampering,
Fig.~\ref{fig:art-inter-safety-in-a} shows $\arti{P}.\iin[{\tt foo}]$
with edges $(p,O_4)$ and $(O_4, O_N)$ removed from
Fig.~\ref{fig:i_in_foo}.
When {\tt regenInter} uses such a tampered ART to analyze {\tt foo} and
reaches the call-site $s_{\ref{ln:moti-ex-foo-rec}}$, it computes
$in_{\tt foo}$ at Line~\ref{ln:art-inter-safety-project-in} of {\tt
checkForInSafety} as shown in Fig.~\ref{fig:art-inter-safety-in-b};
we rediscover the previously removed edge $(p, O_4)$.
This causes the assertion on Line~\ref{ln:art-inter-safety-in-subsumes} to
fail, since $\arti{P}.\iin[{\tt foo}]$ does not encode this information.
In other words, the tampered $\arti{P}.\iin[{\tt foo}]$ does not reflect
the IN-summary of {\tt foo}, at this call-site, hence unsafe to use.

{\bf Establishing Safety of M$_{\bf out}$-Invariants}.
We start with an intuition.
Recall from Section~\ref{ss:art-inter-design} that an M$_{\bf out}$-Invariant encodes the fixed-point context-insensitive OUT-summary of a recursive procedure $M_{rec}$.
This means that for a recursive procedure $M_{rec}$, $\iout[M_{rec}]$ can be used to compute sound fixed-point OUT values of {\em any} of its call-sites.
In contrast, consider a non-conservatively tampered (see Section~\ref{ss:art-intra-safety}) \arti{P}, where the tampering is performed with respect to the $\iout[M_{rec}]$ entry.
In such a case, using $\arti{P}.\iout[M_{rec}]$ (at Line~\ref{ln:art-inter-rec-out},
Fig.~\ref{fig:art-inter-regen-algo}) to compute the OUT value of a
recursive call-site $s$ (in a procedure $M$) will result in the generation of points-to information that does not correspond to the fixed-point context-insensitive points-to information of $M$.
The technique to establish the safety of $\arti{P}.\iout[M_{rec}]$ also makes use of the property of fixed-point values discussed in Section~\ref{ss:art-intra-safety}.
Thus, if $\arti{P}.\iout[M_{rec}]$ represents the fixed-point OUT-summary of $M_{rec}$ obtained after analyzing $M_{rec}$ using its context-insensitive IN-summary in the producer, then processing $M_{rec}$ using the same IN-summary ($\arti{P}.\iin[M_{rec}]$) again in the consumer should not change the OUT-summary of $M_{rec}$.
We use this intuition to define the {\tt checkForOutSafety} method shown
in Fig.~\ref{fig:art-inter-safety-algo}.

For each recursive procedure $t$,
when the OUT-summary is regenerated by {\tt regenInter}  (when the {\tt
Exit} node of $t$ is analyzed by {\tt regenInter}), {\tt
checkForOutSafety} asserts that the regenerated OUT-summary of $t$ matches $\arti{P}.\iout[t]$ (Line~\ref{ln:art-inter-safety-out-subsumes}).
This assertion is important as for any recursive procedure $t$,
a recursive call-site (see Section~\ref{s:back}) invoking 
$t$ must have been processed by {\tt regenInter} before processing the {\tt
Exit} node of $t$, and 
{\tt regenInter} must have used $\arti{P}.\iout[t]$ as the
required OUT-summary of $t$ (at Line~\ref{ln:art-inter-rec-out},
Fig.~\ref{fig:art-inter-regen-algo}), at that call-site. 

If the assertion succeeds, it means that the OUT-summary of $t$
regenerated by {\tt regenInter} matches the fixed-point
context-insensitive OUT-summary of $t$.
Further, this OUT-summary of the procedure $t$ can be used to compute the fixed-point OUT points-to 
information at all of its call-sites.
If the assertion fails, it means that $\arti{P}.\iout[t]$ does not reflect the context-insensitive OUT-summary of $t$.

Similar to the discussion about tampering  via deletion of an ART entry
for the IN-summary,
consider the case where the OUT-summary of a recursive procedure $M_{rec}$ has been
deleted as part of tampering.
In such a scenario, on encountering a recursive call-site $s$ invoking
$M_{rec}$,
{\tt regenInter} sets a default value for 
$\arti{P}.\iout{M_{rec}}$ (= $\arti{P}.\iin{M_{rec}}$), 
at Line~\ref{ln:art-inter-recursive} in
Fig.~\ref{fig:art-inter-regen-algo};
not explicitly shown.
Thus, in case the fixed-point context-insensitive IN-summary of $M$ differs from this default value
then our above discussed procedure will detect the tampering.
And in case they do not differ, then it implies that despite the
tampering,
the ART instance is safe to use; in Section~\ref{s:opt}, we will use this
observation to optimize the encoding of ART.

\begin{figure}[t]
	\centering
	\hfill
	\begin{subfigure}[b]{0.30\textwidth}
	  \centering
	  \includegraphics[width=\textwidth]{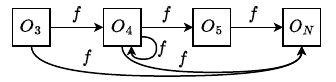}
	  \caption{$\arti{P}.\iout[{\tt foo}]$ tampered by removing edge $(O_5, O_4)$}
	\label{fig:art-inter-safety-out-a}
	\end{subfigure}%
	\hfill
	\hfill
	\begin{subfigure}[b]{0.34\textwidth}
	  \includegraphics[width=\textwidth]{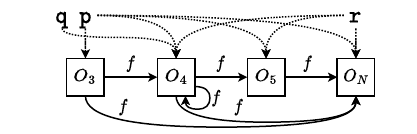}
		\caption{OUT[$s_{\ref{ln:moti-ex-foo-rec}}$] obtained with tampered $\arti{P}.\iout[{\tt foo}]$}
	\label{fig:art-inter-safety-out-b}
	\end{subfigure}%
	\hfill
	\hfill
	\begin{subfigure}[b]{0.30\textwidth}
  	\includegraphics[width=\textwidth]{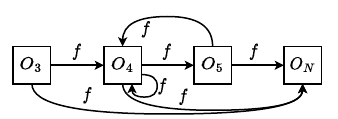}
	\caption{OUT-summary of {\tt foo} computed by {\tt regenInter} ($ \neq \arti{P}.\iout[{\tt foo}]$)}
	\label{fig:art-inter-safety-out-c}
	\end{subfigure}%
	\hfill

	\caption{
		Safety Verification of M$_{\bf out}$-Invariants for
		program in Fig.~\ref{fig:rec-motivating}.
	}

\label{fig:art-inter-safety-out}
\end{figure}

{\bf Example.} 
Consider the ART instance shown in Fig.~\ref{fig:art-inter-proc} for 
the program $P$ shown in Fig.~\ref{fig:rec-motivating}.
As an instance of non-conservative tampering,
Fig.~\ref{fig:art-inter-safety-out-a} shows $\arti{P}.\iout[{\tt foo}]$
with the edge $(O_5,O_4)$ removed from
Fig.~\ref{fig:i_out_foo}.
When {\tt regenInter} uses such a tampered ART 
at the call-site $s_{\ref{ln:moti-ex-foo-rec}}$ (that recursively invokes {\tt foo}), the computed OUT[$s_{\ref{ln:moti-ex-foo-rec}}$] is shown in Fig.~\ref{fig:art-inter-safety-out-b}.
When the {\tt Exit} node of {\tt foo} is eventually processed by {\tt regenInter} and the OUT-summary of {\tt foo} is computed (shown in Fig.~\ref{fig:art-inter-safety-out-c}), we observe that the edge $(O_5, O_4)$ is rediscovered.
This causes the assertion at Line~\ref{ln:art-inter-out-check}
(Fig.~\ref{fig:art-inter-safety-algo}) to fail, since $\arti{P}.\iout[{\tt foo}]$ does not encode this information.
In other words, the tampered $\arti{P}.\iout[{\tt foo}]$ does not reflect the fixed-point context-insensitive OUT-summary of {\tt foo}, and is hence unsafe to use.
It is interesting to note that the edge $(O_5, O_4)$ is missing in
OUT$[s_{\ref{ln:moti-ex-foo-rec}}]$, which means that it does not
correspond to fixed-point points-to information at $s_{\ref{ln:moti-ex-foo-rec}}$.

%% file: algos/art-inter-safety.tex
\SetKwProg{Fn}{Function}{}{end}
\SetKwProg{macro}{Macro}{}{end}
\SetKwFunction{regen}{regenIntra}
\SetKwFunction{nE}{notEmpty}
\SetKwFunction{pop}{pop}
\SetKwFunction{head}{head}
\SetKwFunction{stmts}{statements}
\SetKwFunction{topo}{topologically sorted list of EBBs of}%
\SetKwFunction{cf}{statements of B in control-flow order}%
\SetKwFunction{islead}{is Leader of key EBB}%
\SetKwFunction{getART}{getART}%
\SetKwFunction{get}{get}%
\SetKwFunction{pred}{predOUT}%
\SetKw{assert}{assert}
\SetKw{And}{and}
\SetKw{Or}{or}
\DontPrintSemicolon
\begin{algorithm}[H]
	\small
	\Fn({\em // at call-site $s$, check that \arti{P} subsumes the  IN flow to the target callee $T$})
	{\checkIn{s, T, \arti{P}, \mbox{\rm OUT}}} {
	\text{\rm IN}$[s] \gets \bigsqcap\limits_{{p \in \textsf{preds}(s)}}^{} \text{\rm OUT}[p]$\;
	$in_T \gets \projectin(\text{\rm IN}[s],T)$\;
	\label{ln:art-inter-safety-project-in}
    \assert $\arti{P}.\iin[T] \subsumes in_T$\;
	\label{ln:art-inter-safety-in-subsumes}
	}
\BlankLine
\BlankLine
\BlankLine
\BlankLine
\Fn{\checkOut{M, \mbox{\rm OUT}}} { 
   {
     \assert $\arti{P}.\iout[M]$ = $\out{M}$\;\label{ln:art-inter-out-check} 
	\label{ln:art-inter-safety-out-subsumes}
 }
 }
\BlankLine
\BlankLine
	\caption{Checking Safety of Inter-procedural ART}
	\label{fig:art-inter-safety-algo}
\end{algorithm}

%% file: art-correctness.tex
\subsection{Correctness}
\label{ss:art-correctness}

We now present a correctness argument of ART in the following context: 
the consumer has been given an \artwork{}, from which 
the consumer has to either regenerate sound points-to analysis results, or report a
safety violation.
The consumer uses the algorithm in Fig.~\ref{fig:art-inter-regen-algo} to
regenerate the points-to analysis results from a given \artwork{}.
Without the loss of generality, we also assume that the transfer functions
used by the consumer are monotonic in nature (monotonically increases).

%

Informally, the correctness argument can be summarized in three points:
(i) if there is no tampering then consumer will regenerate results of same
precision as the producer, 
(ii) if the tampering is non-conservative then the consumer will detect
the tampering, and
(iii) if the tampering is conservative, then the consumer will regenerate
a sound, yet an over-approximation of the non-tampered result.
The following two lemmas and corollary formalize the same.

\begin{lemma}
Given a sound iterative-data-flow points-to analysis $A$, 
if $R_{prod}$ represents the sound points-to analysis results (for $A$)
encoded by the producer in the \artwork{} $\Psi$,
then given $\Psi$, the consumer will 
(i) successfully regenerate points-to analysis results
(no safety violation error)
(ii)  and these results will match $R_{prod}$, if
the consumer uses the same lattice and transfer functions as the producer.
\label{lem:lemma1}
\end{lemma}

\extendedAbsract{}{
\begin{proof}
(Proof Sketch.)\\
Since $\Psi$ encodes the results of $R_{prod}$ (which is given to be sound), 
the entries in $\Psi$ encode the sound fixed point points-to results of
all the {\em critical} elements: loop-invariants,
in-summaries of all the functions, and out-summaries of recursive functions.
We will prove the lemma by contradiction.

Say the consumer either 
reported a safety violation error, or
generated results $R'$ that does not
match $R_{prod}$.
We will proceed by doing a case analysis.

{\em Consumer reported a safety violation error.} It implies that the
consumer encountered a violation at the loop-header (loop-invariant
violation), or at the function call (in-summary violation), or at the
exit point of the function (out-summary violation).
This implies that the consumer has discovered some additional information
that was not part of the fixed-point solution - a contradiction.

{\em Consumer generated results $R'\not= R_{prod}$.}
Note: we (re)generate $R'$ from $\Psi$, and the process of regeneration 
exactly matches that of the original points-to analysis generation scheme
(in the producer), except at the three critical elements.
Further, the information at the critical elements is directly taken from
the that of $R_{prod}$.
This implies that the $R'$ must match $R_{prod}$. A contradiction.

Hence proved.

\end{proof}

Note that the precision of the results obtained by a consumer of ART
will naturally depend on the analysis used by the producer. 
}

\begin{lemma}
Given a sound iterative-data-flow points-to analysis $A$, and a program $P$,
if $\Psi$ is a non-conservatively tampered points-to analysis results
(for the analysis $A$ and program $P$), 
then the consumer will detect the violation of the safety.
\label{lem:lemma2}
\end{lemma}

\extendedAbsract{}{
\begin{proof}
(Proof Sketch.)\\

Proof by contradiction: say $\Psi$ is a non-conservatively tampered points-to analysis results
(for the analysis $A$ and program $P$), 
and the consumer did not detect the violation of the safety.

By definition of non-conservative tampering, there exist entries (one ore
more) in $\Psi$ that do not match the fixed-point values for those entries.
It implies that the consumer must find new entries (otherwise, $\Psi$
corresponds to  a fixed-point result and does not correspond to a
non-conservatively tampered \artwork) at one of the critical
entries (could be in the tampered entries or untampered ones). 
It implies that the consumer will find a new value and thus flag a
violation of safety - A contradiction.

\end{proof}
}

\begin{corollary}
Given a sound iterative-data-flow points-to analysis $A$, 
an \artwork{} $\Psi$, and a program $P$, our proposed scheme will
infer a fixed-point solution for the transfer functions in $A$, if
$\Psi$ is safe for $P$ with respect to $A$, and
declare $\Psi$ to be unsafe, otherwise.
\end{corollary}
\extendedAbsract{}{
\begin{proof}
Proof follows from Lemmas~\ref{lem:lemma1} and~\ref{lem:lemma2}.
\end{proof}
}

\extendedAbsract{
See the technical report~\cite{art-arxiv} for a proof sketch, and  
Section~\ref{l:discuss} for a discussion on the possible scenario that
the consumer and producer use different points-to analysis schemes.
}{}.

%% file: opt.tex
\section{Optimizations}
\label{s:opt}

While the techniques discussed in Section~\ref{s:art} ensure that ART
carries only information that is necessary and sufficient for efficient
regeneration of the original sounds points-to analysis, we have identified
four opportunities to improve the efficacy of ART (in terms of reducing the
storage size without compromising on the precision).
We discuss these optimizations below.\\
$\bullet$	{\em Encoding of Loop-Invariants.}
For loops whose body does not contain any statements that affect the heap
(for example, in the case of arithmetic loops),
the fixed-point OUT-value of the loop-header (and all statements
in the loop-body) will be equal to the IN-value of the loop-header.
Thus, the emitted ART instance may be optimized to not carry
loop-invariants for such loop-headers.
During the regeneration phase, if any loop-header $i$
has no entry for $\iloop[i]$ in the given ART instance,
the consumer will simply use the IN-value of the
loop-header as $\iloop[i]$.  
We can easily identify such loops by iterating over the
loop-body and checking for the absence of reference type instructions.
\\
$\bullet$	{\em Encoding of M$_{in}$-Invariants.}
Consider a procedure $M$, called from a set $\mathcal{S}$ of call-sites in
a program $P$.
Say, for each $s\in \mathcal{S}$, 
{\tt project-in}(IN[$s$], $M$) is identical, where
IN[$s$] represents the fixed-point IN value for $s$.
That is, {\tt project-in}(IN[$s$], $M$) is equal to the context-insensitive
IN-summary of $M$.
Some of the common scenarios where we encounter such procedures include,
procedures invoked only once in a program,
static procedures with no arguments, and so on.
For such procedures,
ART can be optimized by avoiding the overhead of carrying the respective
M$_{in}$-Invariants,
and the consumer will use the default value as discussed in
Section~\ref{ss:art-inter-safety}, where we discussed how we handle
tampering by deletion of M$_{in}$-Invariant entries. \\
$\bullet$	{\em Encoding of M$_{out}$-Invariants.}
Consider a recursive procedure $M_{rec}$, in a program $P$.
If the context-insensitive OUT-summary of $M_{rec}$ does not differ from
its context-insensitive IN-summary, then 
for all such procedures,
ART can be optimized by avoiding the overhead of carrying the
M$_{out}$-Invariants,
and the consumer will use the default value as discussed in
Section~\ref{ss:art-inter-safety}, where we discussed how we handle
tampering by deletion of M$_{out}$-Invariant entries. \\
$\bullet$	{\em Avoiding Duplicate Information.}
	When two ART entries refer to the same points-to information, we
	can avoid transmitting duplicate information by transmitting just the unique value and
	using references to the unique entry, wherever required.

%% file: discuss.tex
\section{Discussion}
\label{l:discuss}

In this section, we discuss some interesting features and observations in
the context of ART.

{\bf Handling Tampered \artwork{}.}
\extendedAbsract{
For simplicity, we assume that if a consumer deduces that the input
\artwork{} is tampered with (and unsafe to use, see
Sections~\ref{ss:art-intra-safety}), discards
all the regenerated points-to analysis results and continues
the JIT compilation or analysis like it would, in the absence of our technique.
Such a deduction could be because of tampered \artwork{}, or the consumer
using a points-to analysis whose flow functions do not ``match'' that of the
producer.
}{
The producer (the static java compiler like {\tt javac},  which can have a
static analyzer component, or a static
analyzer like Soot~\cite{soot}), will take a Java
application as input and emit the \artwork{} generated by the static
analyzer, along with the bytecode.
In case the consumer deduces that the input \artwork{} is tampered with (and
hence unsafe to use, see Sections~\ref{ss:art-intra-safety}
and~\ref{ss:art-inter-safety}) then there are two possible paths it can
take: (i)~it can throw an error  and abort the JIT
compilation or analysis being performed by the consumer, or~(ii) give a warning, discard
all the regenerated points-to analysis results and continue
the JIT compilation or analysis like it would in the absence of our technique.
For simplicity, we stick to the option~(ii).
One can also envisage some more strategies between the two paths;
such an exploration is beyond the scope of our current manuscript, and we
leave it as an interesting future work.
Note that 
a consumer may deduce the \artwork{} to be tampered because of an actual
tampered \artwork{}, or the consumer
using a points-to analysis whose flow functions do not ``match'' that of the
producer.
}
\artwork{} may also be conservatively tampered in a way that it now encodes 
{\em overly} imprecise results.
As discussed in Section~\ref{ss:art-intra-safety}, 
ART will admit such results, as they are still sound.
A consumer may decide to use some heuristics to ignore
such possible highly imprecise results.
For example, one such heuristic can be a tunable parameter that sets
a ``usability threshold'', in the form of 
the size of any points-to set, or
an upper limit on the maximum size of 
the encoded \artwork. 
We leave the exploration of such heuristics as a future work.

{\bf No {\rm \artwork{}} transmitted to the consumer.}
A consumer (JIT compiler) capable of using \artwork{} may get a program, without any \artwork{} accompanying it. 
In such a scenario, 
the consumer can easily recognize the complete absence of \artwork{} and not 
attempt the regeneration process at all. 
Instead, the consumer continues as it would in the absence of our technique.


{\bf Dynamic features of Java and ART.}
Java allows the programs to change during program execution using features
like dynamic class loading (DCL) and hot code replace (HCR). 
In such a scenario, we have to analyze the \artwork{} accompanying the
newly loaded code, obtain the points-to information for the newly loaded code,
and establish the overall safety of the \artwork{} of the whole program. 
Efficiently maintaining points-to results in the presence of such dynamic
features is an interesting future work.


{\bf Consumer and producer using a different analysis.}
The regeneration techniques proposed in Sections~\ref{ss:art-intra-regen}
and~\ref{ss:art-inter-regen} assumed that the consumer and the producer
use the same analysis (that is, the transfer functions and lattice).
In the event that the consumer does not know the details about the analysis used by the producer and instead uses an arbitrary points-to analysis, we make the following two observations on the outcome (i) if the 
constraints used by the consumer during regeneration are more precise than the constraints used by the producer to generate \artwork{}, then our technique will regenerate sound points-to analysis for the program; (ii) if the 
constraints used by the consumer are less precise (that is, more
conservative) than the constraints used by the producer, then our
technique may identify it as a case of tampering, since the assertions in
Figures~\ref{fig:art-intra-safety-algo}
and~\ref{fig:art-inter-safety-algo} may fail.
This is expected, as the \artwork{} generated by the
producer may not include the additional information introduced by the more conservative analysis performed by the consumer.

{\bf Threat model.}
Given a program $P$ and an \artwork{} $X$, 
there are two activities that can lead to $X$ being considered tampered with: $X$ has been tampered with, or $P$ has been tampered with. This leads to three scenarios that we consider as part of the threat model:
(i) $X$ is tampered with, but $P$ is not.
(ii) $P$ is tampered with, but $X$ is not.
(iii) Both $X$ and $P$ are tampered with.
All these scenarios can be represented by a single scenario that $X$ is
tampered with respect to $P$, which is the scenario addressed in this manuscript.

{\bf Encoding of \artwork{}.}
We use a simple scheme to encode the \artwork{}.
For each type of information (as discussed in
Section~\ref{ss:art-inter-design}) encoded in
\artwork{}, we emit the corresponding points-to graph (structure discussed
in Section~\ref{s:back}).
We represent local variables by using stack slot indices.
We represent each abstract object as a tuple containing the method
and the program location where the object was allocated.

{\bf Handling Composability.}
Analysis of large real world Java applications with libraries presents two interesting directions of work related to composability. 
(1) Owing to the large sizes and resulting scalability issues with
analysing real world Java applications with libraries, researchers have
proposed analyzing them separately in a modular way~\cite{Cousot:2002}, and
these modular results have been composed\cite{Madhavan:2015,
Calcagno:2011, LattnerLenharthAdve07, rinard, Choi:OOPSLA99,%
10.1145/320384.320400} 
 to realize the analysis results for the whole program. Note that this
composition may involve expensive computation.
Our proposed scheme can be extended in a way the producer sends the ARTworks
for the modular results and the consumer first composes them and then uses the
proposed scheme to validate the composed results.
 The ARTwork for the modular results would have to be extended to maintain the
summary information of all the arguments (including the receiver) at all the
call-sites, for all the possible calls to the unavailable methods.
An important challenge that needs to be addressed in this space is that the
composition at the consumer may involve expensive computations and hence may
not be suitable where the consumer cannot afford to pay a high cost for
obtaining+verifying the analysis results.
(2) The runtime libraries may not be available (and different than the libraries present) during static analysis: \citet{pye} present a scheme in which the application and the runtime libraries are analyzed independently to produce partial analysis results (without being conservative). These two partial analysis results can be combined during the runtime, without losing precision. Extending the idea of our proposed ART framework to partial analysis results and combining them safely would be quite interesting, but beyond the scope of this paper.



%% file: impl.tex
\section{Implementation and Evaluation}
\label{s:impl}
\label{ss:impl}

We have implemented our proposed scheme of ART for Java programs in 
three parts: 
(i) the producer component as an extension to the Soot bytecode
optimization framework~\cite{soot}, and
(ii)~a consumer component (termed \regenpta{}) in the form of a Soot-based static analysis that attempts
to obtain points-to analysis results for a Java program without paying the cost of fixed-point computations, and
(iii)~a second consumer component  in the Eclipse OpenJ9 JIT
compiler~\cite{openj9}, which currently neither performs
nor can afford to perform precise, fixed-point based, points-to analysis due to the performance considerations.
The producer uses an extension to VASCO~\cite{vasco} to obtain flow-sensitive, context-insensitive points-to
information and generate the \artwork{} for each input application.

As is the common practice, 
we use the popular TamiFlex~\cite{tamiflex} tool for resolving reflective calls.
However, note that our proposed scheme is not restricted by the presence of
reflection: 
given any points-to analysis approach (augmented by Tamiflex or
not), 
our scheme can be used to generate the corresponding \artwork{} (by the
producer) and
regenerate the matching points-to information (by the consumer).

%% file: eval.tex
\label{s:eval}

\input{benchmark-table}

To experimentally evaluate the proposed ART scheme,
similar to the many prior works~\cite{pye,ors,10464900},
we used DaCapo and SPECjvm suites.
We chose 13 benchmarks: (i) \textsc{sunflow,
lusearch, luindex \text{\rm and} avrora} from the DaCapo 9.12-MR1-bach
suite~\cite{dacapo}; (ii) \textsc{antlr, fop \text{\rm and} pmd} from the DaCapo
2006-10-MR2 suite; and (iii) \textsc{compress, sparse, sor, fft,
montecarlo, \text{\rm and} lu} from SPECjvm2008~\cite{specjvm}.
The rest of the excluded benchmarks could not be statically analyzed --
either by Soot or by TamiFlex%
\footnote{Our prototype implementation makes use of Soot, along with the {\tt play-in} and {\tt play-out} 
agents of Tamiflex, the combination of which has documented issues with large applications~\cite{issue1,issue2,issue3,issue4,issue5}. 
To compensate, where possible, we have instead used corresponding benchmarks from an older version of the benchmark suite.}.
Our evaluation was performed on a Dell Precision 7920 server, which is a
2.3GHz Intel(R) Xeon(R) Gold 5218 CPU system with 64GB of main memory, running Ubuntu 20.04.1 LTS.
Fig.~\ref{fig:benchmark-tab} shows some of the static characteristics
about the benchmarks used for evaluation;
Column~3 shows the size of the portion of each benchmark analyzed by
Soot (varied between 0.47 to 2.0 MB) and Column~4 shows the number of statically analyzed methods (varied between 466 to 2103) in each benchmark.

We present an evaluation to empirically establish the following 
six research questions:
(RQ1) How does the regeneration technique compare to the complete analysis in terms of time and precision?
(RQ2) Is the proposed scheme able to detect tampering of the generated \artwork{}?
(RQ3) What is the cost of regenerating flow-sensitive, context-insensitive analysis during JIT compilation?
(RQ4) What is the storage overhead of \artwork{}?
(RQ5) What is the effect of the optimizations proposed in Section~\ref{s:opt} on the size of \artwork{}?
(RQ6) What performance benefits can be realized by regenerating analysis results using \artwork{}?

\subsection{Evaluation of the Consumer Components}
\label{sss:eval-jitc}
\begin{figure}[t]
	\begin{subfigure}{0.49\textwidth}
	\small
	\centering
  	\includegraphics[width=\textwidth]{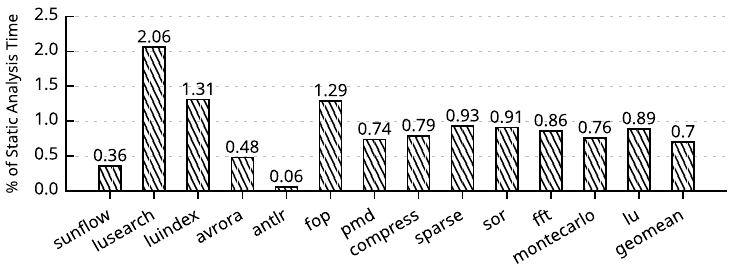}
	\caption{Consumer: \regenpta{}}
	\label{fig:soot-regen-vs-static-time}
\end{subfigure}
\begin{subfigure}{0.49\textwidth}
	\small
	\centering
  	\includegraphics[width=\textwidth]{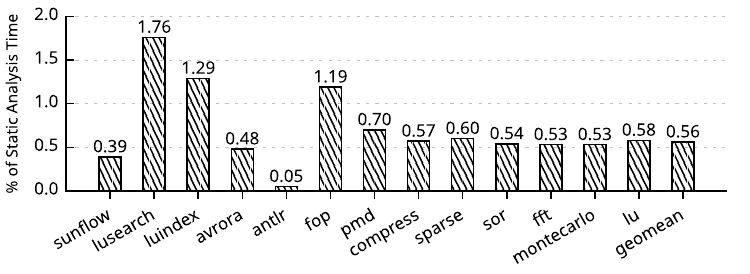}
	\caption{Consumer: OpenJ9 JIT compiler}
	\label{fig:jitc-vs-static-time}
\end{subfigure}
\caption{Time for regeneration in the two consumers as compared to the time 
	taken by the producer to perform the complete analysis.}
\end{figure}
We study the consumer components of our technique with respect to efficiency and safety of the regeneration process.

\noindent {\bf RQ1.} {\em How does the regeneration technique compare to complete analysis in terms of time and precision?}
To answer this question, we compare our regeneration scheme in the
consumers against the complete analysis of the producer.
Figures~\ref{fig:soot-regen-vs-static-time} and~\ref{fig:jitc-vs-static-time} show the time taken to
regenerate the flow-sensitive, context-insensitive points-to analysis
results by our regeneration technique in \regenpta{} and OpenJ9,
respectively, as a percentage of the time taken by the producer to perform
the complete analysis.
The figure shows that our technique regenerates points-to analysis results by
paying an extremely small fraction of the time taken to perform the
complete analysis by the producer (0.06\% to 2.06\%, geomean 0.70\% for
\regenpta{}; and 0.05\% to 1.76\%, geomean 0.56\% for OpenJ9). Note: since the base compilation times are different, the percentage overheads across the two consumers differs.

As an addition, to show that the points-to analysis results regenerated by
the consumer is of equal precision as the results encoded by ART, we
compared the OUT-summary regenerated by the consumers for each procedure
with the OUT-summary of the same procedure as computed by the producer.
These OUT-summaries were found to match for all the procedures analyzed
for each program, thus showing that the regenerated points-to analysis
results are of equal precision to the complete analysis.

\noindent {\bf RQ2.} {\em Is the proposed scheme able to detect tampering of the generated \artwork{}?}
To empirically study the safety of the regenerated results, we tested the ability of our consumer component to identify \artwork{} that is unsafe to use (that is, non-conservatively tampered).
To do this, we randomly identified 10 different elements of \artwork{} for
each benchmark, and non-conservatively tampered them manually.
This process consisted of two steps (1)~randomly picking a non-trivial element of \artwork{} (i.e., one that isn't
solely composed of {\em summary} points-to information), and (2)~non-conservatively tampering it.
Examples of the performed non-conservative tampering include
replacing an object $O_i$ in the points-to set of a variable or a field, with
another object $O_j$ ($i\neq j$),
reducing the cardinality of a points-to set,
and removing a variable or field from a points-to map.
We found that for all such tampered \artwork{},
the consumer successfully identified the tampering.

\begin{figure}[t]
	\small
	\begin{minipage}{0.49\textwidth}
	\centering
  	\includegraphics[width=\textwidth]{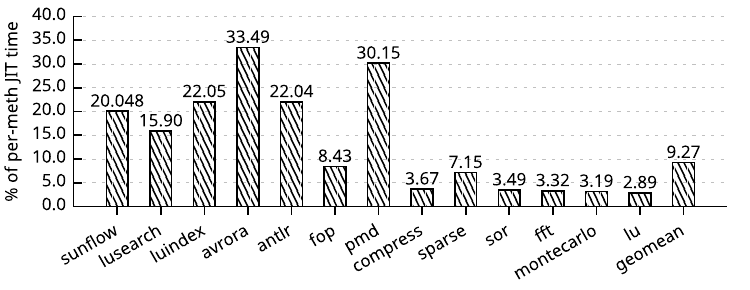}
	\caption{Per-method cost of regeneration in comparison to the
	per-method cost of JIT compilation.}
	\label{fig:per-method-cost-comparison}
\end{minipage}
	\begin{minipage}{0.01\textwidth}
		~
\end{minipage}
	\begin{minipage}{0.49\textwidth}
	\centering
  	\includegraphics[width=\textwidth]{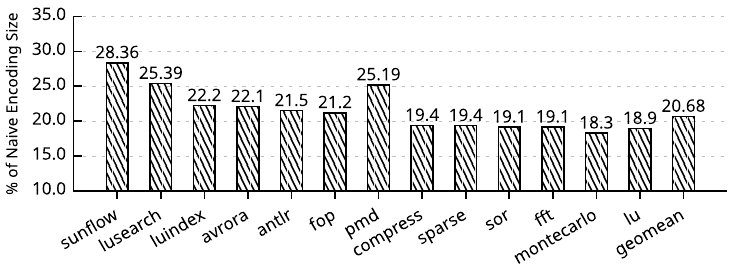}
	\caption{Size of \artwork{} as a percentage of size of naive
	encoding.}
	\label{fig:art-vs-naive}
\end{minipage}
\end{figure}

\noindent {\bf RQ3.} {\em What is the cost of regenerating flow-sensitive, context-insensitive analysis during JIT compilation?}
We have found the regeneration scheme is extremely fast (owing to the
underlying linear-time algorithm).
For example, on average, it takes a couple of milliseconds per
procedure across all the benchmarks.
To understand its performance relative to the existing JIT compilation
time,
Fig.~\ref{fig:per-method-cost-comparison} shows the per-method average time taken by the regeneration
process (includes the time to read the \artwork{} from disk and parse it)
as a percentage of the on-average per-method JIT compilation time. 
We see that we are able to obtain precise points-to information (typically
a very expensive analysis) as a small fraction (15-35\%) of the JIT
compilation time.
We believe that this is reasonable, considering the fact that JIT
compilation itself takes a very small fraction of the total execution time
and the possible utility of the regenerated precise points-to analysis results.
We believe that our proposed technique clearly overcomes the drawbacks
of both the alternatives (of using fast, but highly imprecise heuristics, or precise, but unacceptably
slow analysis, as discussed in Section~\ref{s:intro}), while ensuring the analysis is safe to
use.

\subsection{Evaluation of the Producer Components}
\label{sss:eval-static}
{\bf RQ4.} {\em What is the storage overhead of \artwork{}?}
To study the efficiency of our proposed encoding of \artwork{}, we 
define a {\em naive} encoding as a dump of the OUT values at each
program-point for the whole program, and consider this encoding as a baseline for storage overhead.
Fig.~\ref{fig:art-vs-naive} shows the size of the \artwork{} of each
benchmark, as a percentage of the size of this naive encoding.
The figure shows that the size of \artwork{}, on average, is less than
21\% of the size of the naive encoding; with the maximum size of
\artwork{} being 28.36\% of the naive encoding for any benchmark.

{\bf Note.}
The comparison between the sizes of \artwork{} and naive encoding was performed after compressing the artifacts of each encoding.
Uncompressed, this comparison is observed to favor ART even more.
For example, in the case of \textsc{avrora}, uncompressed ART was 1.7MB in size -- which is only 10\% of the uncompressed naive encoding at 17MB.
Compression tends to make this comparison less dramatic due to the
presence of similar points-to information at different program-points in
the naive encoding, which gets compressed better.
However, using such a naive scheme will naturally lead to 
higher file I/O time because of the handling of larger files.

\begin{figure}[t]
	\small
	\begin{minipage}{0.49\textwidth}
	\centering
  	\includegraphics[width=\textwidth]{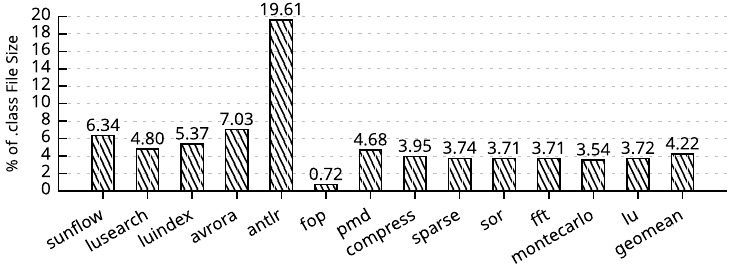}
	\caption{Overhead of \artwork{} in comparison to size of {\tt
	.class} files.}
	\label{fig:art-vs-class}
	\end{minipage}
	\begin{minipage}{0.01\textwidth}
		~
	\end{minipage}
	\begin{minipage}{0.49\textwidth}
	\small
	\centering
  	\includegraphics[width=\textwidth]{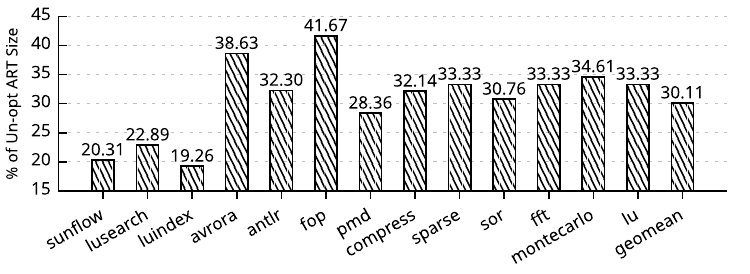}
	\caption{Reduction in size of optimized \artwork{} in comparison
	to size of un-optimized \artwork{}.}
	\label{fig:art-vs-unoptimized-size}
	\end{minipage}
\end{figure}

We also studied the overhead of our encoding of \artwork{} with respect to
the size of the .class files analyzed.
Fig.~\ref{fig:art-vs-class} shows the overhead of \artwork{} as a
percentage of the .class file size.
We find that the overhead is around 4\% of .class files size, on average.

\noindent {\bf RQ5.} {\em What is the effect of the optimizations proposed in Section~\ref{s:opt} on the size of \artwork{}?}
Fig.~\ref{fig:art-vs-unoptimized-size} shows the storage size of the
(optimized) \artwork{} as a percentage of the size of the
\artwork{} obtained without applying the optimizations discussed in Section~\ref{s:opt}.
We observe that the optimized \artwork{} is on average, approximately 70\%
of the size of the unoptimized version (leading to 30\% reduction, on average, in the
size of the \artwork{}).

We have also observed that the regeneration time using the optimized \artwork{} is, on average, approximately 92\% of the regeneration time using
unoptimized \artwork{} (leading to on average 8\% reduction in time taken
for regeneration); we skipped the graph for brevity.
We find the proposed optimizations are effective in reducing the size of
the \artwork{} and reducing the regeneration time.

\subsection{Evaluation of the usability of regenerated analysis results}
\label{sss:eval-inliner}
%
\begin{figure}[t]
	\begin{subfigure}{0.4\textwidth}
	\centering
  	\includegraphics[width=\textwidth]{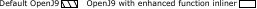}
	\caption*{}
	\end{subfigure}~\\\vspace{-1.5em}
	\begin{subfigure}{0.17\textwidth}
	\centering
  	\includegraphics[width=\textwidth]{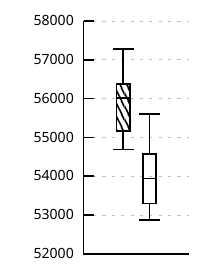}
        \captionsetup{textfont=scriptsize, labelfont=scriptsize}
        \caption{sunflow\\(1862, 1012)}
	\label{fig:boxplot-sunflow}
	\end{subfigure}~\hspace{-1.7em}
	\begin{subfigure}{0.17\textwidth}
	\centering
  	\includegraphics[width=\textwidth]{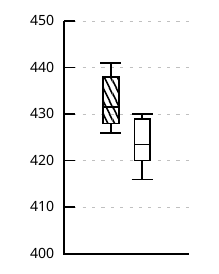}
        \captionsetup{textfont=scriptsize, labelfont=scriptsize}
        \caption{lusearch\\(1860, 1088)}
	\label{fig:boxplot-lusearch}
\end{subfigure}~\hspace{-1.7em}
	\begin{subfigure}{0.17\textwidth}
	\centering
  	\includegraphics[width=\textwidth]{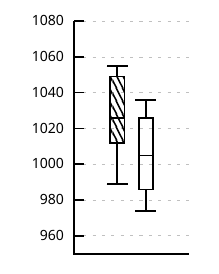}
        \captionsetup{textfont=scriptsize, labelfont=scriptsize}
        \caption{luindex\\(2439, 756)}
	\label{fig:boxplot-luindex}
\end{subfigure}~\hspace{-1.7em}
	\begin{subfigure}{0.17\textwidth}
	\centering
  	\includegraphics[width=\textwidth]{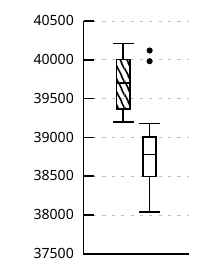}
        \captionsetup{textfont=scriptsize, labelfont=scriptsize}
        \caption{avrora\\(3092, 1396)}
	\label{fig:boxplot-avrora}
\end{subfigure}~\hspace{-1.7em}
	\begin{subfigure}{0.17\textwidth}
	\centering
  	\includegraphics[width=\textwidth]{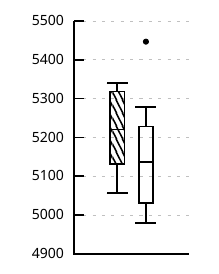}
        \captionsetup{textfont=scriptsize, labelfont=scriptsize}
        \caption{antlr\\(10524, 8295)}
	\label{fig:boxplot-antlr}
\end{subfigure}~\hspace{-1.7em}
	\begin{subfigure}{0.17\textwidth}
	\centering
  	\includegraphics[width=\textwidth]{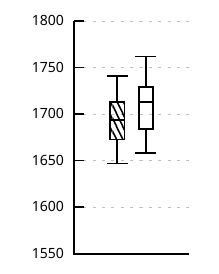}
        \captionsetup{textfont=scriptsize, labelfont=scriptsize}
        \caption{fop\\(598, 0)  }
	\label{fig:boxplot-fop}
\end{subfigure}~\hspace{-1.7em}
	\begin{subfigure}{0.17\textwidth}
	\centering
  	\includegraphics[width=\textwidth]{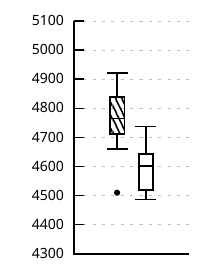}
        \captionsetup{textfont=scriptsize, labelfont=scriptsize}
        \caption{pmd\\(5214, 2579)}
	\label{fig:boxplot-pmd}
\end{subfigure}~\hspace{-1.7em}
	\caption{Comparison of the execution times (in ms) on default OpenJ9 and OpenJ9 with the enhanced inliner.
                The caption subtext shows \#monomorphic-calls-identified
		and \#inlining-guards-avoided.}
	\label{fig:inliner-reduction-exec-time-boxplot}
\end{figure}
\noindent {\bf RQ6.} {\em What performance benefits can be realized by regenerating analysis results using \artwork{}?}
As discussed in the answer to RQ1, 
our first consumer \regenpta{} is able to very efficiently regenerate the points-to analysis results encoded by \artwork{} provided to it
after establishing its safety.
Due to its high efficiency and the modular nature of Soot analyses,
\regenpta{} can be used by any Soot pass (analysis/optimization) that
needs precise points-to analysis results,
without paying the cost of fixed-point computations;
Fig.~\ref{fig:soot-regen-vs-static-time} gives an indication of the
benefits that can be accrued.

Next, to study the impact of efficiently realizing
precise points-to analysis results in the Eclipse OpenJ9 JIT compiler, our second consumer,
we have used 
its function inliner as a client and studied the effect.
Since fixed-point based points-to analysis is considered expensive, OpenJ9 currently uses the following technique to
perform function inlining 
(i) it first uses the information from a runtime profiler
to identify the probabilities of invocation of each potential target of a
virtual call, and then
(ii) if the probability associated with invoking a particular target is sufficiently high, OpenJ9 inlines that particular target
guarded by a runtime check on the type of the receiver of the virtual call.
If the type of the receiver at runtime does not match the type that defined the inlined target, then OpenJ9 falls back to
the original virtual call.
Therefore, in OpenJ9, each inlined call-site takes the form of a
conditional statement -- with the inlined method forming the body of
`then-branch', and the original virtual call the body of the `else-branch'.
In case of {\em monomorphic} calls, as there will always be a singular
invocation target,  the evaluation of the associated guards is redundant.
Removing the corresponding condition-checking code and the else-branch of the
conditional-statement can improve the program performance. We leverage this
intuition to answer RQ6: we used the points-to analysis results regenerated using our technique to identify such monomorphic calls, 
inlined those call-sites without any guards,
and then studied the impact of such an optimization on the overall execution time.
For this experiment, 
we found that SPECjvm offered very few opportunities for applying this
optimization and the gains/overheads were negligible.
Thus, we only show the details for the DaCapo benchmarks.

Fig.~\ref{fig:inliner-reduction-exec-time-boxplot} plots the 
execution times of the benchmarks using
the default OpenJ9, and the OpenJ9 with the enhanced function inliner.
With each benchmark we also mention (in the caption) the number of monomorphic calls
identified  and the number of inlining guards avoided by the enhanced
inliner.
We see that we were able to use the points-to analysis results regenerated by our
technique to identify a large number of monomorphic calls across the
benchmarks evaluated.
Note that for any given program, the number of inlining guards avoided due to this
optimization can be less than the total number of monomorphic
calls identified.
This is because OpenJ9 may not 
find all the monomorphic calls profitable to be inlined (decided by OpenJ9 based on factors like
caller size, callee size, and so on), and hence inlining is not even attempted for such non-profitable calls.

Since the performance of Java applications varies
significantly~\cite{georges}, across different JVM invocations and
iterations, to obtain steady-state performance for
Fig.~\ref{fig:inliner-reduction-exec-time-boxplot}, we used the following
evaluation methodology: for each benchmark, we conducted 30 runs, where each run consisted of 99 warmup iterations, and the following iteration was used to measure the execution time.
The runs for each benchmark were interleaved to account for systemic bias and impact due to different JVM invocations.
As can be seen, the enhanced OpenJ9 is able to reduce execution times for all benchmarks
where a non-zero number of inlining guards were avoided. 
Considering that this improvement is on top of the plethora of other
optimizations that OpenJ9 already has, 
we argue that the gains are significant.

In case of \textsc{fop},
we found that though the enhanced inliner did not reduce any guards,
we did not observe much degradation.
This attests to the extremely low overheads of regeneration 
as discussed in the context of RQ3.


{\em Overall summary.}
Our study shows that the points-to analysis results regenerated using \artwork{} are
not only usable in practice, but can also result 
in measurable performance improvements.

%% file: benchmark-table.tex
\begin{figure}
	\small
\centering
\begin{tabular}{|l|l|c|c|cc|cc|}
\hline
\multicolumn{1}{|c|}{\multirow{2}{*}{ }} & \multicolumn{1}{c|}{\multirow{2}{*}{\textbf{Benchmark}}} & \multirow{2}{*}{\textbf{\begin{tabular}[c]{@{}c@{}}.class size\\ (MB)\end{tabular}}} & \multirow{2}{*}{\textbf{\begin{tabular}[c]{@{}c@{}}\#analyzed\\ methods\end{tabular}}} &  \multicolumn{2}{c|}{\textbf{Artifact Size (KB)}} & \multicolumn{2}{c|}{\textbf{Overhead (\%)}} \\ \cline{5-8} 
\multicolumn{1}{|c|}{} & \multicolumn{1}{c|}{} &  &  & \multicolumn{1}{c|}{\textbf{naive}} & \textbf{ART} & \multicolumn{1}{c|}{\textbf{naive}} & \textbf{ART} \\ \hline
1. & sunflow & 1.2 & 908 &  \multicolumn{1}{c|}{275} & 78 & \multicolumn{1}{c|}{22.38} & 6.34 \\ 
2. & lusearch & 1.6 & 975 &  \multicolumn{1}{c|}{396} & 88 & \multicolumn{1}{c|}{24.17} & 5.37 \\
3. & luindex & 1.3 & 1280 &  \multicolumn{1}{c|}{252} & 64 & \multicolumn{1}{c|}{18.9} & 4.81 \\ 
4. & avrora & 1.5 & 2022 &  \multicolumn{1}{c|}{489} & 108 & \multicolumn{1}{c|}{31.8} & 7.03 \\ \hline
5. & antlr & 1.2 & 1324 &  \multicolumn{1}{c|}{1119} & 241 & \multicolumn{1}{c|}{91.06} & 19.61 \\ 
6. & fop & 1.9 & 377 &  \multicolumn{1}{c|}{66} & 14 & \multicolumn{1}{c|}{3.39} & 0.71 \\ 
7. & pmd & 2.0 & 2103 &  \multicolumn{1}{c|}{381} & 96 & \multicolumn{1}{c|}{18.60} & 4.68 \\ \hline
8. & compress & 0.47 & 466 &  \multicolumn{1}{c|}{98} & 19 & \multicolumn{1}{c|}{20.40} & 3.96 \\ 
9. & sparse & 0.47 & 480 &  \multicolumn{1}{c|}{93} & 18 & \multicolumn{1}{c|}{19.32} & 3.74 \\ 
10. & sor & 0.47 & 480 &  \multicolumn{1}{c|}{94} & 18 & \multicolumn{1}{c|}{19.53} & 3.74 \\
11. & fft & 0.47 & 485 &  \multicolumn{1}{c|}{94} & 18 & \multicolumn{1}{c|}{19.40} & 3.72 \\ 
12. & montecarlo & 0.47 & 485 &  \multicolumn{1}{c|}{93} & 17 & \multicolumn{1}{c|}{19.40} & 3.54 \\ 
13. & lu & 0.47 & 487 &  \multicolumn{1}{c|}{95} & 18 & \multicolumn{1}{c|}{19.66} & 3.72 \\ \hline
\multicolumn{1}{|c}{}& geomean & 0.85 & 728 &  \multicolumn{1}{c|}{157.89} & 33.72 & \multicolumn{1}{c|}{18.02} & 4.22 \\ \hline
\end{tabular}
	\caption{Static Details of the benchmarks used, storage overheads
	for ART and whole analysis results.}
	\label{fig:benchmark-tab}
\end{figure}

%% file: related.tex
\section{Related Work}
\label{s:related}
~

{\bf Staged Analysis.}
There have been prior works that have attempted to reduce JIT compilation overheads due to expensive analyses by leveraging the multi-stage nature of Java compilation.
For example~\citet{ali_2014,serrano,10.5555/902440} have proposed schemes to perform
expensive whole-program analyses/optimizations statically for Java, which
are then leveraged at runtime to obtain complete analysis/optimized code.
Similarly, the PYE framework~\cite{pye} performs points-to analysis in a static analyzer and sends along partial summaries representing the analysis to the JIT compiler,
where it is augmented with information available at runtime to make it more precise.
We note that all these schemes are affected by the same challenges safety and transmission costs discussed in Section~\ref{s:intro}, thereby severely impacting their practical usage.
Our proposed technique of ART addresses both these issues and can be the first step towards making staged analysis practical.



%

{\bf Points-to Analysis.}
There have been many papers on points-to/alias analysis, depicting various dimensions of analysis such as flow-sensitivity, context-sensitivity, path-sensitivity, and so on~\cite{pye,dietrich,
rinard,tan,10.1145/320384.320400}, various avenues for speeding up \cite{ThakurNandivada19,ThakurNandivada20,RamaKomondoorSharma18}, 
and performing optimization specific analysis~\cite{pye,NandaSinha09,loginov08}, and so on.
For our implementation, we have extended the analysis implemented in VASCO~\cite{vasco} to build our static analyzer.
Many recent static analysis tools~\cite{10.1145/3428228, 10.1145/3450492, 10.1145/3622832, 10.1145/3527311, 10.1145/3276511} employ SPARK~\cite{10.1007/3-540-36579-6_12} (which is flow- and context-insensitive, to ensure scalability) as their points-to analysis solution.
Deploying ART in these tools can improve their precision 
by offering flow-sensitive points-to analysis, with little additional cost.
Over the years, there has been a rich body of work championing flow-sensitive, context-insensitive 
points-to analyses~\cite{10.1145/320384.320400, 10.1007/978-3-642-31057-7_29, 6993367, 934854}.
Rather than serve as a replacement for these techniques, ART in fact promotes their adoption by ameliorating the concern of scalability in tools that wish to employ them.

{\bf Declarative points-to analyses. }
Datalog-based points-to analyses (like Doop~\cite{10.1145/1640089.1640108}) are convenient due to the
declarative nature of their analysis rules.
While this paper discusses ART in the context of IDFA-based points-to analysis,
ART isn't restricted by an IDFA-based producer.
We believe that ART can be deployed even in systems where the producer 
employs a declarative points-to
analysis (see prior work~\cite{10.1145/3213846.3213860,10.1145/2491956.2462191,10.1145/1926385.1926390}) 
provided the consumer possesses an equivalent set of transfer functions to 
apply during the regeneration process.

{\bf Interactive Proof Systems.}
A keen reader will note that ART is similar, in spirit, to interactive proof
systems like zero-knowledge protocols~\cite{10.1145/22145.22178} (ZKPs).
While an interesting parallel, we note that they differ fundamentally in both 
domain and purpose of application. 
ZKPs are employed in cryptography to convey an awareness (i.e., a proof)
of knowledge without actually giving away the knowledge, an interaction motivated
by privacy.
In contrast,
ART is not motivated by privacy, but with
efficiently sharing points-to information, along with a proof of its correctness.


%

%% file: concl.tex
\section{Conclusion and Future work}
\label{s:concl}
This article proposes a scheme called ART that helps 
Java compilers (both static and JIT) and program analysis tools -- termed {\em consumers} -- 
obtain highly precise (flow-sensitive, context-insensitive) points-to
analysis results in an efficient and safe manner; where such analysis
results were hitherto difficult to attain due to performance and safety concerns.
Using ART, a consumer can obtain very precise points-to analysis results
without prohibitively impacting  
compilation (or analysis) time or compromising on the safety of the analysis results.
We show, via a detailed evaluation, that ART enables the generation of precise
and trusted analysis results in consumer systems in a minute
fraction of the time (<1\%, on average) required to perform the complete analysis,
while adding a very small space-overhead (around 4\%, on average) to the bytecode.
It would be interesting to extend the proposed ideas to support other
dimensions of analyses (for example, call-site-based context-sensitivity,
object-sensitivity, path-sensitivity, and so on) and  the nuances of
other non-Java like languages.
For example, supporting context-sensitivity necessitates an efficient encoding 
in ART to uniquely identify each calling context;
and languages like C and C++ may need interesting augmentations to ART to support features
like pointer arithmetic, casting and function pointers.
We leave such explorations as future work.

\begin{acks}
    This research was partially supported by  \grantsponsor{IBM}{IBM}{} CAS projects
    \grantnum{IBM}{1101} and~\grantnum{IBM}{1156}. We also thank Manas
    Thakur from IIT Bombay for the fruitful discussions during the early
    stages of this work. 
\end{acks}